\documentclass[10pt,journal,compsoc]{IEEEtran}
\ifCLASSOPTIONcompsoc
\usepackage[nocompress]{cite}
\else
  \usepackage{cite}
\fi
\usepackage{algorithm}
\usepackage{algpseudocode}
\usepackage{epsfig}
\usepackage{graphicx}
\usepackage{amsfonts}
\usepackage{comment}
\usepackage{caption}
\usepackage{color}
\usepackage{amssymb}
\usepackage{amsmath}
\usepackage{amsthm}
\usepackage{multirow}
\usepackage{enumitem}
\usepackage{etoolbox}

\newtheorem{lemma}{Lemma}
\AfterEndEnvironment{lemma}{\noindent\ignorespaces}

\newtheorem{corollary}{Corollary}
\newtheorem{definition}{Definition}[section]
\AfterEndEnvironment{definition}{\noindent\ignorespaces}
\newtheorem{example}{Example}[section]
\AfterEndEnvironment{example}{\noindent\ignorespaces}

\ifCLASSINFOpdf
\else
\fi
\begin{document}
\title{QoS constrained Large Scale Web Service Composition using Abstraction Refinement}

\author{Soumi~Chattopadhyay,~\IEEEmembership{Student Member,~IEEE,} and
        Ansuman~Banerjee,~\IEEEmembership{Member,~IEEE}

\IEEEcompsocitemizethanks{\IEEEcompsocthanksitem 
This work has been submitted to the IEEE for possible publication. Copyright may be transferred without notice, after which this version may no longer be accessible.
}

}

\makeatletter
\long\def\@IEEEtitleabstractindextextbox#1{\parbox{0.922\textwidth}{#1}}
\makeatother

\IEEEtitleabstractindextext{
\begin{abstract}
Efficient service composition in real time, while satisfying desirable Quality of Service (QoS) guarantees for the composite solution has 
always been one of the topmost research challenges in the domain of service computing. On one hand, optimal QoS aware service composition algorithms, 
that come with the promise of solution optimality, are inherently compute intensive, and therefore, often fail to generate the optimal solution 
in real time for large scale web services. On the other hand,  heuristic solutions that have the ability to generate solutions fast and 
handle large and complex service spaces, settle for sub-optimal solution quality. The problem of balancing the trade-off between compute efficiency 
and optimality in service composition has alluded researchers for several decades, and several   
proposals for taming the scale and complexity of web service composition have been proposed in literature. In this paper, we present 
a new perspective towards this trade-off in service composition based on abstraction refinement, that can be seamlessly integrated on 
top of any off-the-shelf service composition method to tackle the space complexity, thereby, making it more time and space efficient. 
Instead of considering 
services individually during composition, we propose a set of abstractions and corresponding refinements to form service groups based on functional characteristics. 
The composition and QoS satisfying solution construction steps are carried out in the abstract service space. 
Our abstraction refinement methods give a significant speed-up compared to traditional 
composition techniques, since we end up exploring a substantially smaller space on average. Experimental results on 
benchmarks show the efficiency of our proposed mechanism in terms of time and the 
number of services considered for building the QoS satisfying composite solution.
\end{abstract}

}

\maketitle

\IEEEdisplaynontitleabstractindextext
\IEEEpeerreviewmaketitle
\ifCLASSOPTIONcaptionsoff
  \newpage
\fi

\begin{figure*}[!ht]
\centering
\includegraphics[width=\linewidth]{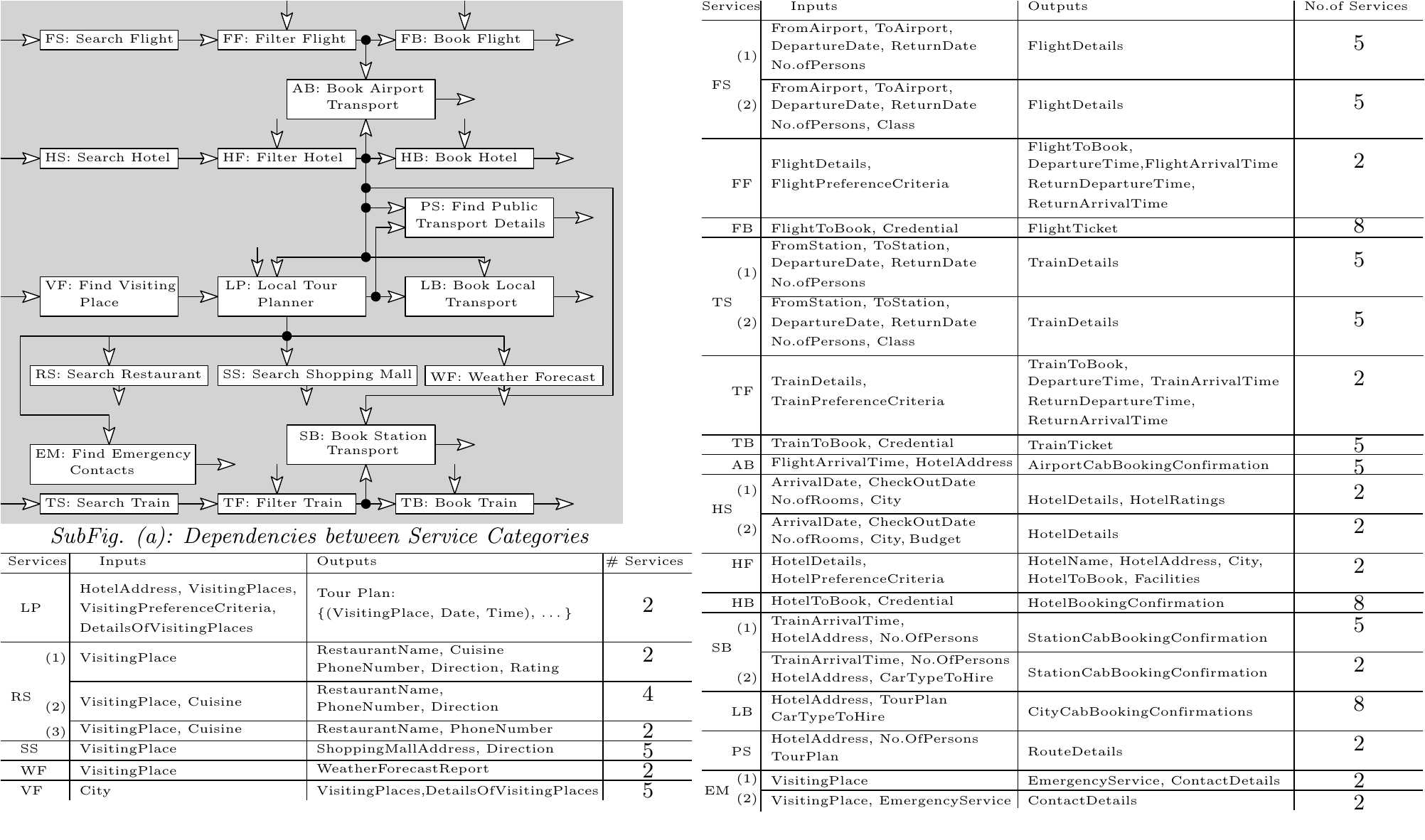}
\caption{Service repository}
\label{fig:defInSR}
\end{figure*}
\IEEEraisesectionheading{\section{Introduction}\label{sec:intro}}
\noindent
In recent times, web services have become the most powerful gateway to massive repositories of information and applications to serve user queries. 
A web service is a software component that performs a specific task and is characterized by its functional attributes (inputs / outputs) and non functional Quality of Service (QoS) attributes like response time, throughput, reliability, availability, 
compliances etc. In many cases, it is not possible to serve a user specified query with a single web service, for which{\textcolor{blue}{,} composition is required 
to invoke multiple web services in a specific order. During service composition, on one side, it is 
important to ensure that functional dependencies are met, i.e., the input-output dependencies of the services are honored. On the other hand, it is also important to ensure the overall non functional 
requirements, i.e., overall QoS values of the solution \cite{7226855} and satisfy constraints on them as applicable. 
Scalability of composition algorithms for real time query response has always been a major element of concern \cite{milanovic2004current,lecue2009towards}. With tremendous increase in the scale of web services \cite{7027801}, large scale web service composition \cite{rodriguez2015hybrid} is becoming an increasingly important challenge. 

A large body of literature in service composition deals with methods for 
computing the optimal composition \cite{schuller2012cost, aiello2009optimal, papadias2003optimal} satisfying QoS constraints as applicable. These methods, though having the promise of optimality of the composite solution, often fail to scale to large web service repositories for real time response at query time. Service composition approaches based on A*\cite{6009375}, constraint satisfiability\cite{1357986} and graph search\cite{chattopadhyay2015scalable, DBLPChattopadhyayB16}, fail to deliver the expected performance, due to the inherent complexities of the methods they use to navigate and search for the optimal solution in the composed state space. Comparable advances have also been made in the area of heuristic approaches towards service composition~\cite{chattopadhyay2015scalable}, wherein, the solution optimality is compromised to make the solution generation fast, scalable and efficient at query time. Indeed, balancing the trade-off between optimality and efficiency has always been a foremost research challenge with several widely varying proposals~\cite{6009422, alrifai2009combining, 5552793} in literature.

In this work, we examine the problem of QoS constrained query time service composition from a completely different perspective, and propose a technique that can be seamlessly integrated on top of any off-the-shelf service composition method to tackle the space complexity, thereby, making it more time and space efficient. 
Our technique is based on the idea of {\em{abstraction refinement}} \cite{clarke2000counterexample}, which was originally developed for handling large state spaces in the context of formal verification.
Our proposal has two major steps. As a first step, we aim to expedite the solution computation time of the underlying service composition method.
Instead of considering 
services individually during composition, we propose a set of abstractions and corresponding refinements to form service groups based on functional characteristics and the composition is carried out in the abstract service space. Our approach has the guarantee of always being able to generate a QoS satisfying solution, if one exists.
Abstraction reduces the search space significantly, and thus speeds up the composition and solution generation steps. While this can expedite the solution construction step to a great extent, this also entails a possibility that it may fail to generate any solution satisfying a given set of QoS constraints, though the individual services allow a valid solution. Therefore, as a second step, we propose to refine an abstraction to generate the composite solution with desired QoS values. A QoS satisfying solution, if one exists, can be constructed from the abstraction with refinement. In the worst case (which is rare in our experiments), our approach may end up exploring the complete composition graph constructed on individual services, and thereby never miss a solution if one exists. In general, a solution \textcolor{black}{satisfying all the QoS constraints} can be obtained efficiently from the abstract graph. \textcolor{black}{Abstraction is done at preprocessing / design time, whereas, the composition is done at query time. Though the abstraction step does not affect run-time composition performance, the refinement step (if needed), being done at query time, does add up a non-negligible overhead to the composition process. However, as can be seen from our experiments, the underlying service composition method is still much faster in the average case, when compared with its performance without the abstraction and refinement steps. }

We perform extensive experiments on the popular benchmarks ICEBE-2005~\cite{icebe2005} and WSC-2009~\cite{bansal2009wsc} to demonstrate the power of abstraction. In each case, as evident from the experiments detailed in Section~\ref{sec:result}, our method adds orders of magnitude scalability, when considered on top of three recent~\cite{xia2013web,DBLPChattopadhyayB16,6009375} service composition methods. 

\section{Motivating Example}\label{sec:overview}
\begin{figure}[!ht]
\centering
\includegraphics[width=\linewidth]{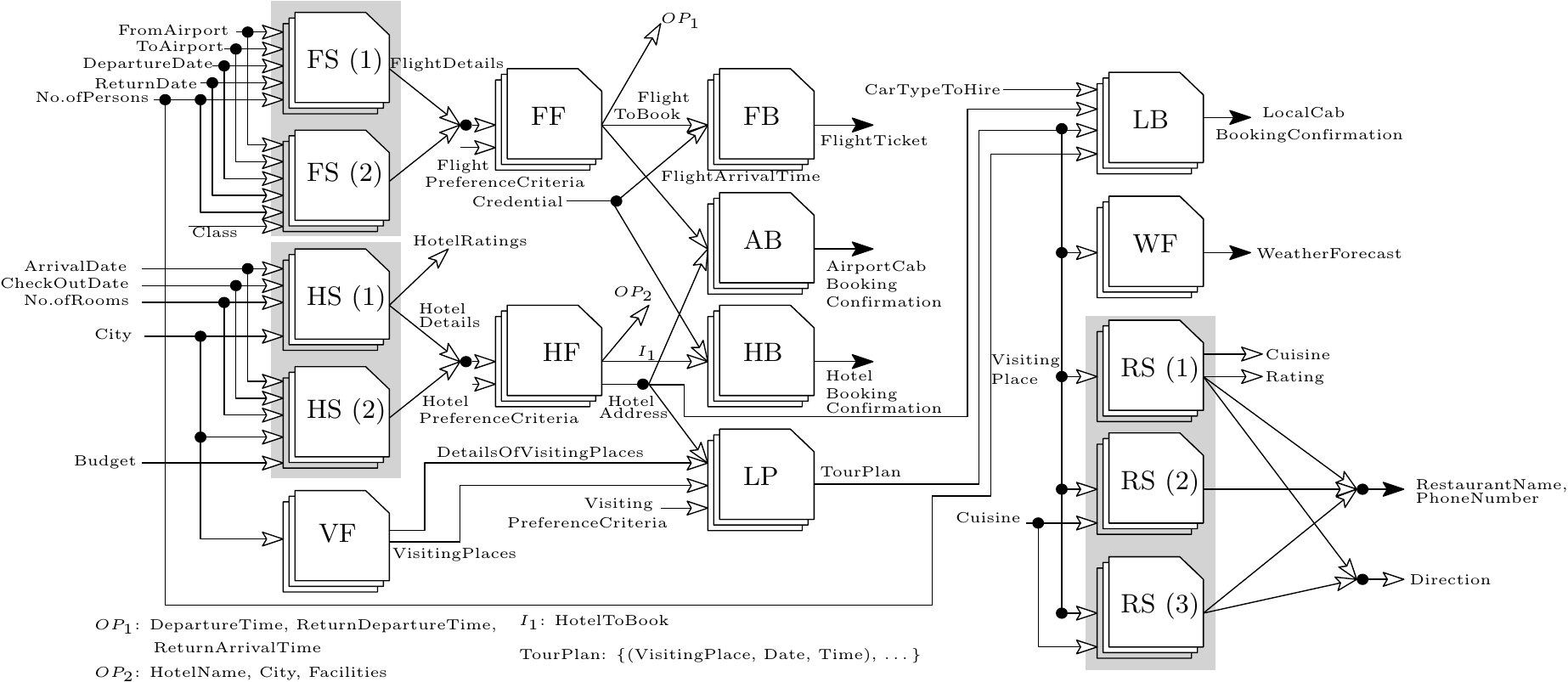}
\caption{Service dependency graph to serve ${\cal{Q}}$}
\label{fig:dependencyInSR}
\end{figure}
\vspace{-0.2cm}
\noindent
We use an example description of a service repository for tour planning, to illustrate our methods throughout the paper.
The service repository contains 99 services. The services are broadly classified into 19 different service categories depending on the 
specific task each service caters to, as shown in Figure \ref{fig:defInSR}.
SubFigure {\em{(a)}} of Figure \ref{fig:defInSR} shows the dependencies
between the service categories. Each rectangular box of the subFigure 
represents a service category and a directed edge between
two service categories represents an input-output dependency, i.e. some / all outputs generated by the source task go as inputs into the destination task. 
The input-output sets of the services for each category are shown in the $2^{nd}$ and the $3^{rd}$ columns of the table of Figure \ref{fig:defInSR}. 
For a specific category, there are multiple services which are able to fulfill the service task, and these are shown as sub-categories in the $1^{st}$ column. 
The input-output set of two services in a service category can be different, in which case they are placed in an appropriate sub-category. 
Consider the flight search (FS) service category consisting of 10 services. As shown in Figure \ref{fig:defInSR}, there are two sub-categories, 
$FS^{(1)}$ and $FS^{(2)}$, differing in their input sets. Out of the 10 services for $FS$, 5 services have the same input set and are in sub-category $FS^{(1)}$, 
whereas, the other 5 services have a different input set and are in sub-category $FS^{(2)}$. The outputs of both the sub-categories are same in this case. 
A similar clustering based on both inputs and outputs can be seen for the service task $EM$. It may be noted that services 
in the same category / sub-category may have different QoS values. For the sake of simplicity, we consider here only two QoS parameters, namely, the response time 
and invocation cost. Each service, therefore, is associated with values of these two parameters. 

Consider a query ${\cal{Q}}$ with inputs {\em{\{FromAirport, ToAirport, DepartureDate, ReturnDate, No.ofPersons, Class, FlightPreferenceCriteria, 
Credential, ArrivalDate, CheckOutDate, No.ofRooms, City, Budget, HotelPreferenceCriteria, VisitingPreferenceCriteria, Cuisine\}}} and desired output 
{\em{\{FlightTicket, HotelBookingConfirmation, AirportCabBookingConfirmation, CityCabBookingConfirmations, WeatherForecastReport, RestaurantName, 
PhoneNumber\}}}. 
Consider constraints on maximum response time and invocation cost for the composite service as 1 second and \$10 respectively.
The objective is to serve ${\cal{Q}}$, while satisfying the QoS constraints.
In classical service composition methods, the input-output functional service dependencies are represented by a directed graph, 
usually referred to as the functional dependency graph~\cite{xia2013web}. The composition solution corresponding to a service query is either 
a path or a subgraph of the dependency graph, depending on the functional parameter~\cite{xia2013web} and the nature of the services being 
composed \cite{Alrifai:2012:HAE:2180861.2180864}. The size of the dependency graph depends on the number of services in the service repository 
and their inputs and outputs. A service repository may contain a large number of web services. In this case, construction of the dependency graph 
incurs a lot of time and space. 
Considering the above example, the functional dependency graph for serving ${\cal{Q}}$ involving 64 services is shown in Figure \ref{fig:dependencyInSR}. 
Each block in Figure \ref{fig:dependencyInSR} represents multiple services having identical set of inputs and outputs.  

Managing the complexity of the dependency graph and the solution extraction step thereafter, has been the foremost 
challenge for composition. Optimal service composition methods based on ILP~\cite{schuller2012cost} or sub-optimal heuristics~\cite{alrifai2009combining} both have their own limitations in handling such graphs at the web scale, since they traverse the concrete service space to generate a solution. While optimal methods encode the entire dependency graph as a set of constraints, heuristics employ smart traversal mechanisms to tame the space complexity. Our perspective in contrast is to work with any of these methods but on a reduced abstract space, created from the concrete service repository, and refined at query time. We show through our experiments that the composition and QoS constraint satisfying solution construction steps are more efficient when considered on the abstracted reduced search space using any of the methods discussed above. In this paper, we propose several such abstractions that can expedite any composition method to a significant extent. 

\section{Detailed Methodology}\label{sec:method}
\noindent
The service composition problem considered here is same as in classical services work~\cite{7226855}. {\textcolor{black}{The inputs to the problem are:}
\begin{itemize}
 \item A set of web services $S = \{{\cal{S}}_1, {\cal{S}}_2, \ldots, {\cal{S}}_n\}$
 \item For each web service ${\cal{S}}_i \in S$, a set of inputs ${\cal{S}}_i^{ip}$ and a set of outputs ${\cal{S}}_i^{op}$
 \item For each web service ${\cal{S}}_i \in S$, a set of QoS parameters (e.g., Response time, Reliability, Invocation cost etc) {\textcolor{black}{represented as a tuple}}
 \item A set of concepts~\cite{DBLP:journals/tweb/EshuisLM16} (optional) that capture information about semantically equivalent service descriptions
 \item A query ${\cal{Q}}$, specified by a set of inputs ${\cal{Q}}^{ip}$ and a set of requested outputs ${\cal{Q}}^{op}$
 \item An optional set $\cal{C}$ of QoS constraints {\textcolor{black}{represented by a tuple}} on the composite solution providing 
	bounds on the worst case values
\end{itemize}

\noindent
The objective of classical QoS constrained service composition is to serve ${\cal{Q}}$ by suitable composition of the services in 
$\{{\cal{S}}_1, \ldots, {\cal{S}}_n\}$ preserving functional dependencies and satisfying all QoS constraints. The complexity of finding a QoS satisfying optimal solution depends on the QoS parameter under consideration. On one hand, for QoS parameters like response time and throughput, the complexity is polynomial in the number of services~\cite{yan2015anytime}. On the other hand, for QoS parameters like reliability, availability, reputation, invocation cost, the problem is NP-hard \cite{yan2015anytime}.

As discussed earlier, 
our motivation in this work is to create a framework that can seamlessly fit on top of any service composition approach and expedite 
its performance. 
In the following discussion, we explain how our abstraction methods leverage on the functional characteristics of the services. 
Abstractions are done in a preprocessing phase, before composition and query arrival. In this paper, we consider four levels of 
abstraction. In each level, one or multiple services are abstracted by a new service. The main objective of abstraction is to 
reduce the number of services to be considered for query processing during composition.

\vspace{-0.3cm}

\subsection{Abstraction of equivalent services}
\noindent
Before we discuss the details of the abstraction, we begin with a few definitions.

\begin{definition}{\em [\bf Input (Output) Equivalence:]}\label{def:inputEqui}
 Two services ${\cal{S}}_1, {\cal{S}}_2$ are input (output) equivalent, expressed as ${\cal{S}}_1 \simeq_i {\cal{S}}_2$ (${\cal{S}}_1 \simeq_o {\cal{S}}_2$) 
 if their input (output) sets are identical.
 \hfill$\blacksquare$
\end{definition}

\noindent
We now illustrate the notion of a {\em concept} and how it helps in finding identical inputs and outputs between services, even when the names do not match exactly.
An input (output) $i_1$ ($o_1$) of a service ${\cal{S}}_1$ is said to be identical to an input (output) $i_2$ ($o_2$) of a service ${\cal{S}}_2$, if both 
of them semantically refer to the same concept. Consider two flight search services ${\cal{S}}_1$ and ${\cal{S}}_2$. The input set of ${\cal{S}}_1$ is \{from, 
to, departureDate, noOfPersons\} and the input set of ${\cal{S}}_2$ is \{fromAirport, toAirport, departureDate, noOfPersons\}. The inputs of ${\cal{S}}_1$ and 
${\cal{S}}_2$ are considered identical since both of them refer to the same {\em concept}, i.e., departure airport and arrival airport respectively. A set of 
such concepts capture additional semantic information between service descriptions and accompany any service repository~\cite{DBLP:journals/tweb/EshuisLM16}.

We now present a few results that help us {\textcolor{black}{to}} build the foundations of our abstraction and the guarantees they offer. All proofs are available in the appendix.
\begin{lemma}
 The binary relation $``\simeq_i"$ defined over a set of services is an equivalence relation.
 \hfill$\blacksquare$
\end{lemma}



\noindent
On similar lines, $``\simeq_o"$ is an equivalence relation as well.

\begin{definition}{\em [\bf Equivalent Services:]}
 Two services ${\cal{S}}_1, {\cal{S}}_2$ are equivalent (${\cal{S}}_1 \simeq {\cal{S}}_2$), 
 if they are input equivalent, ${\cal{S}}_1 \simeq_i {\cal{S}}_2$, as well as output equivalent, 
 ${\cal{S}}_1 \simeq_o {\cal{S}}_2$.
 \hfill$\blacksquare$
\end{definition}

\noindent
The following holds with a similar reasoning.

\begin{lemma}\label{lemma:equivalence}
 The binary relation $``\simeq"$ defined over a set of services is an equivalence relation.
 \hfill$\blacksquare$
\end{lemma}

\noindent
The notion of equivalence is purely syntactic, in other words, two equivalent services can have different QoS parameter values. In the first level 
of abstraction, we abstract equivalent services by a new service. The services in the service repository are divided into equivalence classes 
($``\simeq"$ is an equivalence relation). The set of all equivalence classes of $S$ forms a partition of $S$. Every service of $S$ thus belongs to 
one and only one equivalence class. Each equivalence class is then represented by an abstract service. This reduces the number of services in the 
dependency graph considered at query time. Therefore, computing the solution to a query becomes easier. Evidently, the QoS values of 
the composite solution may be affected as a result of this abstraction step.

Each equivalence class of services {{$\{{\cal{S}}_{i1}, {\cal{S}}_{i2}, \ldots, {\cal{S}}_{ik}\} \subseteq S$}} is abstracted by a 
new representative service ${\cal{S}}^{(1)}_{i}$. The inputs and outputs of ${\cal{S}}^{(1)}_{i}$ are same as ${\cal{S}}_{ij}$.
Consider {{$S^{(1)} = \{{\cal{S}}^{(1)}_1, {\cal{S}}^{(1)}_2, \ldots, {\cal{S}}^{(1)}_{n_1}\}$}} be the set of $n_1$ abstract 
services generated from the original set of $n$ services.
Evidently, the set of services corresponding to ${\cal{S}}^{(1)}_i$ and ${\cal{S}}^{(1)}_j$ are mutually exclusive for $i \neq j$.
Therefore, the following holds.





\begin{lemma}
 The number of services after the first level of abstraction is less than or equal to the total number of services in the 
 service repository, i.e., $n_1 \le n$. \hfill$\blacksquare$
\end{lemma}


\noindent
Algorithm \ref{algo:equivalent} details the abstraction procedure. Step \ref{step6} is the key step of Algorithm \ref{algo:equivalent}.
In this step, for each service ${\cal{S}}_i \in S$, all other services in the repository
are examined to identify whether they are equivalent to ${\cal{S}}_i$. The complexity of this step is $O(n)$
and thereby, the overall complexity of the algorithm is $O(n^2)$, where $n$ is the number of services in the original service repository. 

\begin{algorithm}
\scriptsize
  \caption{Abstraction of Equivalent Services}
  \begin{algorithmic}[1]
   \State {Input: $S = \{{\cal{S}}_1, {\cal{S}}_2, \ldots, {\cal{S}}_n\}$}
   \State{Output: Abstract Service set $S^{(1)} = \{{\cal{S}}^{(1)}_1, {\cal{S}}^{(1)}_2, \ldots, {\cal{S}}^{(1)}_{n_1}\}$}
   \State $S' = S$;
   \Repeat
    \State  Remove an arbitrary service ${\cal{S}}_i$ from $S'$;
    \State ${\cal{U}}$ = services equivalent to ${\cal{S}}_i$;\label{step6}
    \State Construct a service ${\cal{S}}^{(1)}_i$ corresponding to ${\cal{U}} \cup \{{\cal{S}}_i\}$;
    \State Inputs of ${\cal{S}}^{(1)}_i$ = inputs of ${\cal{S}}_i$;
    \State Outputs of ${\cal{S}}^{(1)}_i$ = outputs of ${\cal{S}}_i$;
    \State Add ${\cal{S}}^{(1)}_i$ to $S^{(1)}$;
    \State $S' \leftarrow S' \smallsetminus ({\cal{U}} \cup \{{\cal{S}}_i\})$; \Comment{$``\smallsetminus"$ represents set minus operator}
   \Until{$S'$ is not empty}
  \end{algorithmic}
  \label{algo:equivalent}
\end{algorithm}

\begin{example}
\textcolor{black}{
 We now {\textcolor{black}{illustrate}} the effect of this abstraction step on our motivating example discussed in Section \ref{sec:overview}. 
 The services in a sub-category have 
 identical set of inputs and outputs  and therefore, they are equivalent. Hence, each sub-category, consisting of multiple services, is abstracted by a new service. 
 The service repository  contains 26 sub-categories in total. Therefore, the number of services after this abstraction step is reduced from 99 to 26. 
 Each block in Figure \ref{fig:dependencyInSR}, represents a single abstract service instead of multiple services.}
 \hfill$\blacksquare$
\end{example}

\noindent
The crux of this abstraction is to merge equivalent services based on input-output equivalence. The challenge now is to assign a 
{\em representative QoS value} to the created abstract service, since services inside an equivalence class may have widely different 
QoS values. For this, we choose for each class, the {\em best representative} service $s$ (defined formally below) and assign the QoS 
values of $s$ to the abstract service. The choice of $s$ for each class is done as explained below.

Consider an abstract service ${\cal{S}}^{(1)}_i \in S^{(1)}$ corresponding to an equivalence class 
${\hat{S}}_{(i)} = \{{\cal{S}}_{i1}$, ${\cal{S}}_{i2}$, $\ldots, {\cal{S}}_{ik}\}$.
If there exists any service ${\cal{S}}_{ij} \in {\hat{S}}_{(i)}$, such that, it has the best values for all the QoS parameters among 
all the services in ${\hat{S}}_{(i)}$, we assign the values of all QoS parameters of ${\cal{S}}_{ij}$ to ${\cal{S}}^{(1)}_i$. 
Otherwise, we choose the best representative one in terms of the QoS parameter values. 
Since the QoS parameters are disparate in nature, we first normalize their values for each service in an equivalence class. 
QoS parameters (QP) like reliability and availability, for which a high value is desirable (i.e., positively monotonic), are normalized as:

{\scriptsize{
\begin{equation}
NV ({\cal{S}}_{ij} (QP)) = 
\begin{cases} 
      1  \text{\indent\indent\indent\indent~~~, if }Max(QP) = Min(QP) \\
      \frac{{\cal{S}}_{ij} (QP) - Min(QP)}{Max(QP) - Min(QP)}\text{, otherwise }
   \end{cases}
\end{equation}}}
\vspace{-0.2cm}

\noindent
where ${\cal{S}}_{ij} (QP)$ is the value of QP for service 
${\cal{S}}_{ij} \in \{{\cal{S}}_{i1}$, ${\cal{S}}_{i2}$, $\ldots, {\cal{S}}_{ik}\}$, $Max(QP)$ and $Min(QP)$ are the 
maximum and minimum values of QP across the set ${\cal{S}}_{i1}$, ${\cal{S}}_{i2}$, $\ldots, {\cal{S}}_{ik}$ respectively.
Similarly, for QoS parameters like response time, invocation cost, total number of invocations for which a low value is desirable (i.e., negatively monotonic), 
the normalization is defined as:

{\scriptsize{
\begin{equation}
NV ({\cal{S}}_{ij} (QP)) = 
\begin{cases} 
      1  \text{\indent\indent\indent\indent~~~, if }Max(QP) = Min(QP) \\
      \frac{Max(QP) - {\cal{S}}_{ij} (QP)}{Max(QP) - Min(QP)} \text{, otherwise }
   \end{cases}
\end{equation}}}
\vspace{-0.2cm}

\noindent
\textcolor{black}{The normalized value (NV) of each QoS parameter (QP) is positively monotonic and lies between 0 to 1.}
Once we normalize the QoS parameters, 
we compute the deviation of the QoS parameter of a service from the best value of the QoS parameter as follows:

{\scriptsize{
\begin{equation}
Deviation({\cal{S}}_{ij} (QP)) = 1 - NV ({\cal{S}}_{ij} (QP))
\end{equation}
}}
\vspace{-0.6cm}

\noindent
We now define the notion of the best representative service.

\begin{definition}{\em [\bf Best Representative Service:]}
 The best representative service of an equivalence class is a service for which the maximum deviation across all QoS parameters
 is minimum among all services in that class.\hfill$\blacksquare$
\end{definition}

\noindent
If multiple services exist for which the maximum deviation is minimum, {\textcolor{black}{we resolve the tie by considering 
the second maximum deviation for each service across all parameters and selecting the service for which
the second maximum deviation is minimum among the tied services.}} We continue the procedure until 
the tie is resolved. Once we select the best representative service $s$, we assign all the values of its QoS parameters to ${\cal{S}}^{(1)}_i$. Algorithm \ref{algo:equivalentQoSAssignment} presents the formal algorithm. 
We illustrate our method on the example below.

\noindent
\begin{example}
 Consider two input-output equivalent services ${\cal{S}}_1$ and ${\cal{S}}_2$ and two QoS parameters, namely, invocation cost and reliability. 
 The QoS values of ${\cal{S}}_1$ and ${\cal{S}}_2$ are (100, 0.96) and (150, 0.98) respectively.
 By our abstraction mechanism, ${\cal{S}}_1, {\cal{S}}_2$ are abstracted by a new service ${\cal{S}}^{(1)}_1$. 
 To assign the value of {\textcolor{black}{the}} QoS parameters of ${\cal{S}}^{(1)}_1$, 
  we compute the $NV$ values as follows.
{{
 $NV ({\cal{S}}_{1} (Invocation Cost)) = 1$, $NV ({\cal{S}}_{1} (Reliability)) = 0.43$, $NV ({\cal{S}}_{2} (Invocation Cost)) = 0.83$, $NV ({\cal{S}}_{2} (Reliability)) = 0.71$. }}
 Consider the maximum and minimum values of invocation cost among all services are 400 and 100 respectively and 
 the same for reliability are 0.99 and 0.93 respectively.
 The maximum deviation for ${\cal{S}}_1$ across the two QoS parameters is $Max((1 - 1), (1 - 0.43)) = 0.57$, 
 the same for ${\cal{S}}_2$ is $Max((1 - 0.83), (1 - 0.71)) = 0.29$.
 We select ${\cal{S}}_2$ and assign the values of the QoS parameters of ${\cal{S}}_2$ to ${\cal{S}}^{(1)}_1$, since for ${\cal{S}}_2$, the maximum deviation across all parameters is minimum. 
\hfill $\blacksquare$
 \end{example}

\begin{algorithm}
\scriptsize
  \caption{Identify representative service}
  \begin{algorithmic}[1]
   \State {Input: ${\hat{S}}_{(i)} = \{{\cal{S}}_{i1}$, ${\cal{S}}_{i2}$, $\ldots, {\cal{S}}_{ik}\}$}\Comment{Set of equivalent services}
   \State{Output: ${\cal{S}}_{il} \in {\hat{S}}_{(i)}$}\Comment{Chosen service for QoS assignment}
   \For {${\cal{S}}_{ij} \in {\hat{S}}_{(i)}$}
    \For {each $QP \in $ Set of QoS parameters}
     \State Compute $NV ({\cal{S}}_{ij} (QP))$;
     \State Compute $Deviation({\cal{S}}_{ij} (QP)) = 1 - NV ({\cal{S}}_{ij} (QP))$;
    \EndFor
    \State $\delta ({\cal{S}}_{ij}) \leftarrow MAX_{QP \in  \text{Set of QoS parameters}}$ ($Deviation({\cal{S}}_{ij} (QP))$);
   \EndFor
   \State Identify the service ${\cal{S}}_{il} \in {\hat{S}}_{(i)}$, such that 
	  $\delta ({\cal{S}}_{il}) = Min_{{\cal{S}}_{ij} \in {\hat{S}}_{(i)}} \delta ({\cal{S}}_{ij})$;\\
	  \Comment{In case of a tie, break the tie as discussed above}
  \end{algorithmic}
  \label{algo:equivalentQoSAssignment}
\end{algorithm}

\noindent
We now discuss the correctness and optimality guarantee of our QoS assignment mechanism. 
Consider a service ${\cal{S}}^{(1)}_i \in S^{(1)}$ corresponding to its equivalence class ${\hat{S}}_{(i)}$.
The QoS assignment to ${\cal{S}}^{(1)}_i$ ensures the following lemma.

\begin{lemma}\label{lemmaOptimalSingle}
 For a single QoS parameter, the QoS value assigned to ${\cal{S}}^{(1)}_i$ is optimal among the set of QoS values in ${\hat{S}}_{(i)}$.  
 \hfill$\blacksquare$
\end{lemma}


 
 \begin{lemma}\label{lemmaOptimalMultiple}
 For multiple QoS parameters, the QoS values assigned to ${\cal{S}}^{(1)}_i$ belong to the Pareto-optimal front \cite{marler2004survey2} 
	constructed over all the QoS values corresponding to ${\hat{S}}_{(i)}$. 
 \hfill$\blacksquare$
\end{lemma}

\subsubsection{Composition with abstract services}
\noindent 
We can {\textcolor{black}{now}} use any service composition method at query time over the abstract service set and construct a composite 
solution that satisfies all QoS constraints, if one exists. 
The major advantage of the first abstraction step is its guarantee of solution quality (more importantly, optimality) preservation for a single QoS parameter case. Thus, after this abstraction step, if an optimal method is used for composition on the abstract set of services, we are guaranteed to be able to generate the same optimal solution that would have been generated by the optimal method, had the composition been carried out in the original un-abstracted service space without abstraction. On a similar note, if a heuristic approach is used, it is guaranteed that a solution of the same quality can be generated over the abstract space. The results are formalized in the discussion below. 

\begin{corollary}\label{corollaryOptimal}
For a single QoS parameter, equivalence abstraction is solution quality preserving.
 \hfill$\blacksquare$
\end{corollary}

\noindent
Corollary \ref{corollaryOptimal} directly follows from Lemma \ref{lemmaOptimalSingle}.
However, for multiple QoS parameters, such a claim does not hold good. Consider the following example.

\begin{example}\label{example:multiOptimalSolution}
 Consider two equivalence classes $\{{\cal{S}}_1, {\cal{S}}_2, {\cal{S}}_3\}$ and $\{{\cal{S}}_4, {\cal{S}}_5, {\cal{S}}_6\}$.
 Consider each service has three QoS parameters{\textcolor{black}{:}} response time, throughput and invocation cost. The values of the QoS parameters 
 for each service are shown as follows: ${\cal{S}}_1: (120, 50, 10), {\cal{S}}_2: (75, 75, 25), {\cal{S}}_3: (125, 100, 20)$, 
 ${\cal{S}}_4: (75, 50, 20), {\cal{S}}_5: (150, 50, 10), {\cal{S}}_6: (175, 50, 10)$. According to our first abstraction strategy,
 $\{{\cal{S}}_1, {\cal{S}}_2, {\cal{S}}_3\}$ is abstracted by ${\cal{S}}^{(1)}_1$ and $\{{\cal{S}}_4, {\cal{S}}_5, {\cal{S}}_6\}$
 is abstracted by ${\cal{S}}^{(1)}_2$. Following our QoS assignment mechanism, ${\cal{S}}_3: (125, 100, 20)$ is chosen as the best 
 representative for the first equivalence class and thus its QoS values are assigned to ${\cal{S}}^{(1)}_1$. Similarly, being
 the best representative of the second class, the QoS values of ${\cal{S}}_5: (150, 50, 10)$ are assigned to 
 ${\cal{S}}^{(1)}_2$. It may be noted, in both the cases, the QoS values of the best representative service belong to the Pareto optimal 
 frontier of their respective equivalence classes. Let us now assume ${\cal{S}}^{(1)}_1$ and ${\cal{S}}^{(1)}_2$ can be composed sequentially.
 The QoS values of the composite service ${\cal{CS}} : {\cal{S}}^{(1)}_1$ followed by ${\cal{S}}^{(1)}_2$ are $(275, 50, 30)$.
 However, the QoS values of ${\cal{CS}}$ in the composition space of original services do not belong to the Pareto optimal solution frontier,
 since, the sequential composition of ${\cal{S}}_1$ and ${\cal{S}}_5${\textcolor{black}{, having QoS values as $(270, 50, 20)$,}} dominates 
 ${\cal{CS}}: (275, 50, 30)$. 
\hfill$\blacksquare$
\end{example}


\noindent
As discussed, for multiple QoS parameters, equivalence abstraction 
is not solution quality preserving.
We now present some results on the guarantees of this abstraction. 

\begin{lemma}{{\textbf{\em{Soundness theorem}}}: }\label{soundAbstraction}
 A QoS constraint satisfying abstract solution is also a valid solution 
 satisfying all QoS constraints on the original set of services. 
 \hfill$\blacksquare$
\end{lemma}
\vspace{-0.5cm}
%


\begin{lemma}{{\textbf{\em{Preservation theorem}}}: }
 We can always construct a solution (in terms of functional dependencies) by composing the abstract services,
 if and only if there exists a solution to a query using the original set of services. 
 \hfill$\blacksquare$
\end{lemma}
\vspace{-0.5cm}
\begin{proof}
 To prove this, we present the following intuitive argument. By accumulating the equivalent services into one abstract service, we essentially remove multiple identical paths 
 (in terms of functional dependencies) from the dependency graph. As noted earlier, each abstract service is functionally equivalent to services in its equivalence class. 
 Consider ${\cal{S}}^{(1)}_i$ be an abstract service and ${\hat{S}}_{(i)}$ be its corresponding equivalence class. Therefore, if any service ${\cal{S}}_{ij} \in {\hat{S}}_{(i)}$ is eventually activated from the query inputs, ${\cal{S}}^{(1)}_i$
 is also activated by the same and thereby, will produce identical set of outputs as produced by ${\cal{S}}_{ij} \in {\hat{S}}_{(i)}$, which 
 leads to producing the query output eventually, if there exists a solution to the query using the original set of services .
\end{proof}

\noindent

\begin{lemma}{{\textbf{\em{No-Loss theorem}}}: }
 All feasible solutions to a query can be generated from the abstract service space. \hfill$\blacksquare$
\end{lemma}
%

\subsubsection{First level refinement}
\noindent
The only problem using abstraction is that for multiple parameters, composition using abstract services may not entail any solution 
{\em{satisfying the QoS constraints}}, 
even if one exists in the original service space, as illustrated below. 

\begin{example}
 Consider the example discussed in Example \ref{example:multiOptimalSolution}. Also consider a query $\cal{Q}$, which requires
 ${\cal{S}}^{(1)}_1$ and ${\cal{S}}^{(1)}_2$ to be composed sequentially. Further, consider the QoS constraints as $(200, 50, 50)$. 
 {\textcolor{black}{The}} solution ${\cal{CS}} : (275, 50, 30)$, returned by the composition algorithm (which works on the abstract service space),
 does not satisfy all the constraints, although, a solution satisfying all constraints exists in the original service space.
 \hfill$\blacksquare$
\end{example}

\noindent
For single QoS aware composition, no refinement is required, since for a single QoS parameter,
equivalence abstraction is solution quality preserving, as stated in Corollary \ref{corollaryOptimal}. 
{\textcolor{black}{However, for multiple parameters, it is necessary.}}
Using refinement, we gradually recover the original services from the abstract services and reconstruct the solution.
We discuss two different refinements below. 

\subsubsection{QoS-based solution refinement}
\noindent 
In this strategy, we modify the QoS values of a solution by improving the QoS assignment to the abstract services, in such a way that the constraints can be satisfied. 
We illustrate this step below.
Consider a solution ${\cal{P}}$ constructed over the set of abstract services and returned by the composition algorithm. Assume ${\cal{P}}$ violates some of the QoS constraints. Each abstract service ${\cal{S}}^{(1)}_i$ in ${\cal{P}}$ corresponds to an equivalence partition ${\hat{S}}_{(i)}$. According to our abstraction algorithm, a representative service is chosen for each 
equivalence partition and its QoS values are assigned to the corresponding abstract service. Our objective here is to 
select another service from the equivalent set, so that the QoS constraints are satisfied. 
We now define two key concepts.


\begin{definition}{\em [\bf Laxity:]}
The laxity ${\cal{L}}$ for a QoS parameter $q$ for a constraint satisfying solution $\cal R$ is the {\em{difference}} between 
the bound on $q$ and the value of $q$ in ${\cal R}$.
 \hfill$\blacksquare$
\end{definition}

\begin{definition}{\em [\bf QoS violation gap:]}
The QoS violation gap ${\cal{V}}$ for a solution ${\cal{P}}$ violating a QoS constraint on a parameter $q$ is the {\em{difference}} 
between the bound on $q$ and the value of $q$ in ${\cal{P}}$.
 \hfill$\blacksquare$
\end{definition}


\noindent
{\textcolor{black}{
Consider a QoS parameter $q$ with a bound $\delta^{(q)}$ and a solution to a query satisfying the QoS constraint with the 
value $\delta'^{(q)}$ of $q$ in the solution. If the solution satisfies the constraint, $\delta'^{(q)}$ must be as good 
as $\delta^{(q)}$. The laxity ${\cal{L}}(\delta^{(q)}, \delta'^{(q)})$ is then mathematically defined as the {\em{difference}} 
between $\delta'^{(q)}$ and $\delta^{(q)}$. 
Similarly, if the solution violates the constraint, $\delta^{(q)}$ must be better than $\delta'^{(q)}$. 
The QoS violation gap ${\cal{V}}(\delta^{(q)}, \delta'^{(q)})$ is then mathematically defined as the {\em{difference}} between 
$\delta'^{(q)}$ and $\delta^{(q)}$.
For different QoS parameters, this {\em{difference}} may be calculated in different ways. 
For example, consider the response time $q$. 
If the response time constraint is satisfied by ${\cal{P}}$, ${\cal{L}}(\delta^{(RT)}, \delta'^{(RT)}) = {\delta}^{(RT)} - {\delta'}^{(RT)}$.
If the response time constraint is violated by ${\cal{P}}$, ${\cal{V}}(\delta^{(RT)}, \delta'^{(RT)}) = {\delta'}^{(RT)} - {\delta}^{(RT)}$. 
Consider another example, where $q$ is reliability.
If the reliability constraint is satisfied by ${\cal{P}}$, ${\cal{L}}(\delta^{(R)}, \delta'^{(R)}) = \frac{{\delta}^{(R)}}{{\delta'}^{(R)}}$. 
In case of reliability constraint violation, ${\cal{V}}(\delta^{(R)}, \delta'^{(R)}) = \frac{{\delta'}^{(R)}}{{\delta}^{(R)}}$.
The laxity and QoS violation gap definitions for other QoS parameters are discussed in Appendix B.
}}
%
%
%



We now discuss the QoS based solution refinement strategy. Consider we have $m$ different QoS parameters with constraints on them. 
Also consider for a particular solution ${\cal{P}}$, out of $m$ 
different QoS constraints, $m_1$ are satisfied (assuming $m_1 < m$). Thereby $(m - m_1)$ QoS constraints are violated. It may be 
noted, the solution ${\cal{P}}$ is constructed on a set of abstract services. Consider ${\cal{C}}_{SAT}$ be the set of QoS parameters that 
are satisfied by ${\cal{P}}$. For each QoS parameter $q \in {\cal{C}}_{SAT}$, we calculate the laxity of $q$. Once we calculate the 
laxity of all QoS parameters in ${\cal{C}}_{SAT}$, we start refining ${\cal{P}}$. Consider $\xi$ be the laxity set of QoS parameters in 
${\cal{C}}_{SAT}$, where $\epsilon_{q_i} \in \xi$ is the laxity of $q_i \in {\cal{C}}_{SAT}$. Our objective is to relax the QoS parameters 
in ${\cal{C}}_{SAT}$ up to their laxity and tighten the rest of the QoS parameters so that the QoS violation gap becomes 0.

\begin{algorithm}
\scriptsize
  \caption{Update QoS parameters}
  \begin{algorithmic}[1]
   \State {Input: ${\cal{S}}^{(1)}_{ij}$, ${\hat{S}}_{(j)}$, $s$, ${\cal{C}}_{SAT}$, $\xi$}
   \State ${\hat{S}}'_{(j)} = \phi$;
   \For {$s' \in {\hat{S}}_{(j)}$}
      \For {$q_i \in {\cal{C}}_{SAT}$}
	\State $val(q_i)\leftarrow$ Relax(value of $q_i$ of $s$, $\epsilon_{q_i}), \epsilon_{q_i} \in \xi$;
      \EndFor
      \If {(the values of all $q_i \in {\cal{C}}_{SAT}$ of $s'$ is better than or equal to value of $q_i$) \indent and 
	   (the values of all $q_i \in ({\cal{C}} \setminus {\cal{C}}_{SAT})$ of $s'$ is better than or equal to
	   the \indent value of $q_i$ of $s$)} 	 \label{line:selection}  
	    \State ${\hat{S}}'_{(j)} = {\hat{S}}'_{(j)} \cup \{s'\}$;
      \EndIf
   \EndFor
   \State $s'\leftarrow$ service from ${\hat{S}}'_{(j)}$ with maximum gain;
   \State Assign QoS parameters of $s'$ to ${\cal{S}}^{(1)}_{ij}$;
  \end{algorithmic}
  \label{algo:sub_first_refine}
\end{algorithm}
 \vspace{-0.6cm}
\begin{algorithm}
\scriptsize
  \caption{Refine QoS parameters}
  \begin{algorithmic}[1]
   \State {Input: ${\hat{S}}^{(1)}_i$, ${\cal{C}}_{SAT}$, $\xi$}
   \For {${\cal{S}}^{(1)}_{ij} \in {\hat{S}}^{(1)}_i$}
    \State Update QoS parameters (${\hat{S}}_{(j)}$, $s$, ${\cal{C}}_{SAT}$, $\xi$);
    \State $\xi \leftarrow$ Revise the laxity of each QoS in ${\cal{C}}_{SAT}$;
    \If {laxity of each QoS in ${\cal{C}}_{SAT}$ is 0}
      \State Break;
    \EndIf
   \EndFor
  \end{algorithmic}
  \label{algo:first_refine}
\end{algorithm}

\noindent
Consider ${\hat{S}}^{(1)}_i = \{{\cal{S}}^{(1)}_{i1}, {\cal{S}}^{(1)}_{i2}, \ldots, {\cal{S}}^{(1)}_{ik}\}$ be the set of 
abstract services in ${\cal{P}}$ which correspond to more than one original service. We attempt to revise the QoS parameters. 
We also consider the set of services ${\hat{S}}_{(j)}$ corresponding to each ${\cal{S}}^{(1)}_{ij} \in {\hat{S}}^{(1)}_i$ and
the service $s \in {\hat{S}}_{(j)}$ whose QoS parameters are initially assigned to ${\cal{S}}^{(1)}_{ij}$. 
Algorithm \ref{algo:sub_first_refine} presents the mechanism for revising the QoS parameters of ${\cal{S}}^{(1)}_{ij}$. 
The essential idea of this algorithm is to select a service $s'$ from ${\hat{S}}_j$ so that we can reduce the QoS violation gap. 
$s'$ is chosen such that the following conditions are satisfied:

\begin{itemize}
 \item For each QoS parameter $q_i \in {\cal{C}}_{SAT}$, the value of $q_i$ of $s'$ is better than or equal to 
	$Relax (\text{value of $q_i$ of }s, \epsilon_{q_i})$, for $\epsilon_{q_i} \in \xi$. The $Relax()$ step relaxes the value of the QoS parameter of $s$ upto a limit. For different QoS parameters, this step is different. For example, in case of response time, we add the laxity of the response 
	time with the response time of $s$. For reliability, we multiply the reliability laxity with the reliability of $s$.
 \item For rest of the QoS parameters, the values of the parameters of $s'$ are at least as good as $s$.
 \item There can be more than one service satisfying the above conditions. In this case, we choose the service
	which provides the maximum gain, defined as:
	
	\noindent
	{\scriptsize
	\begin{equation}\label{eq:gain}
	  gain(s_i) = \sum_{q_j \in ({\cal{C}} \setminus {\cal{C}}_{SAT})}(NV (s'_i (q_j)) - NV (s_i (q_j)))
	\end{equation}}
	where ${\cal{C}}$ is the set of QoS parameters.	
\end{itemize}

\noindent
Algorithm \ref{algo:first_refine} shows the overall refinement procedure. Algorithm \ref{algo:first_refine}
internally calls Algorithm \ref{algo:sub_first_refine} to update the QoS value of an abstract service.
The complexity of Algorithm \ref{algo:sub_first_refine} is $O(|{\hat{S}}_{(j)}|)$, i.e., order of the number 
of equivalent services corresponding to an abstract service. Therefore, the overall complexity of Algorithm 
\ref{algo:first_refine} is the order of the number of abstract services in the solution multiplied by the
number of equivalent services corresponding to each abstract service in the solution. Hence, the worst case 
complexity of the algorithm is $O(n)$; where $n$ is the number of services in the service repository. The worst 
case arises when all the services in the service repository are involved {\textcolor{black}{in computing}} a single solution (e.g., if 
the service repository consists of only the equivalent set of services corresponding to each abstract service 
in the solution). The QoS based solution refinement technique deals with the original services corresponding 
to each abstract service of the solution. Therefore, after the QoS based solution refinement, we end up having the 
solution constructed over the original services.

\subsubsection{Complete refinement}
The QoS parameter based refinement technique discussed above cannot always produce a valid solution, even though there exists one.
The major limitation of this technique is that instead of analyzing the dependency graph, it always proceeds on a 
solution satisfying functional dependencies. 

\begin{example}
Consider three equivalence classes $\{{\cal{S}}_1, {\cal{S}}_2, {\cal{S}}_3\}$, $\{{\cal{S}}_4, {\cal{S}}_5, {\cal{S}}_6\}$ and $\{{\cal{S}}_7, {\cal{S}}_8\}$.
 Also consider, each service has three QoS parameters{\textcolor{black}{:}} response time, throughput and invocation cost. The values of the QoS parameters 
 for each service are shown as follows: ${\cal{S}}_1: (120, 50, 10), {\cal{S}}_2: (75, 75, 25), {\cal{S}}_3: (125, 100, 20)$, 
 ${\cal{S}}_4: (75, 50, 20), {\cal{S}}_5: (150, 50, 10), {\cal{S}}_6: (175, 50, 10), {\cal{S}}_7: (175, 50, 10), 
 {\cal{S}}_8: (50, 45, 15)$. According to our first abstraction strategy,
 $\{{\cal{S}}_1, {\cal{S}}_2, {\cal{S}}_3\}$ is abstracted by ${\cal{S}}^{(1)}_1$,
 $\{{\cal{S}}_4, {\cal{S}}_5, {\cal{S}}_6\}$ is abstracted by ${\cal{S}}^{(1)}_2$ and 
 $\{{\cal{S}}_7, {\cal{S}}_8\}$ is abstracted by ${\cal{S}}^{(1)}_3$. 
 Following our QoS assignment mechanism, ${\cal{S}}_3: (125, 100, 20)$ is chosen as the best representative service for the first equivalence class 
 and thus its QoS values are assigned to ${\cal{S}}^{(1)}_1$. Similarly, being the best representative service of the second and the third equivalence classes, the 
 QoS values of ${\cal{S}}_5: (150, 50, 10)$ and ${\cal{S}}_7: (175, 50, 10)$ are assigned to ${\cal{S}}^{(1)}_2$ and ${\cal{S}}^{(1)}_3$ respectively. 
 Let us now assume ${\cal{S}}^{(1)}_1$ can be composed sequentially with ${\cal{S}}^{(1)}_2$ or ${\cal{S}}^{(1)}_3$.
 The QoS values of the composite service ${\cal{CS}}_1 ({\cal{S}}^{(1)}_1$ followed by ${\cal{S}}^{(1)}_2$) and 
 ${\cal{CS}}_2 ({\cal{S}}^{(1)}_1$ followed by ${\cal{S}}^{(1)}_3$) are $(275, 50, 30)$ and $(300, 50, 30)$ respectively. 
 Consider, for a query ${\cal{Q}}$, either we need to compose ${\cal{S}}^{(1)}_1$ and ${\cal{S}}^{(1)}_2$ or
 ${\cal{S}}^{(1)}_1$ and ${\cal{S}}^{(1)}_3$. Also consider the QoS constraint as $(125, 50, 50)$.
 If the composition solution returns ${\cal{CS}}_1: (275, 50, 30)$ (being the best), the QoS aware solution refinement cannot produce 
 any QoS satisfying solution. 
  \hfill$\blacksquare$
\end{example}

\noindent
For cases as above, we consider our second strategy of complete refinement. 
In this method, the abstract services are replaced by the original services in the dependency graph. We then 
reconstruct the solution from the original dependency graph. It is obvious that complete refinement always generates a QoS constraint satisfying solution.

\subsection{Abstraction based on functional dominance}
\noindent
This abstraction is based on the notion of dominance.

\begin{definition}{\em [\bf Dominant Service:]}
 A service ${\cal{S}}_1$ dominates ${\cal{S}}_2$ (${\cal{S}}_1 \succ {\cal{S}}_2$), if the input set of ${\cal{S}}_2$ 
 is a {\em{superset}} of the input set of ${\cal{S}}_1$ and the output set of ${\cal{S}}_2$ is a {\em{subset}} of 
 the output set of ${\cal{S}}_1$, i.e., ${\cal{S}}_1^{ip} \subseteq {\cal{S}}_2^{ip}$ and ${\cal{S}}_1^{op} \supseteq {\cal{S}}_2^{op}$.
 \hfill$\blacksquare$
\end{definition}

\begin{example}
 Consider two services ${\cal{S}}_1$ and ${\cal{S}}_2$. The input set of ${\cal{S}}_1$ is $\{location\}$ and output set of 
 ${\cal{S}}_1$ is $\{restaurantName, phoneNumber, cuisine, rating\}$, whereas, the input set of ${\cal{S}}_2$ is $\{location, cuisine\}$ and output set of 
 ${\cal{S}}_2$ is $\{restaurantName, phoneNumber\}$. In this case, ${\cal{S}}_1 \succ {\cal{S}}_2$, since ${\cal{S}}_1^{ip} \subset {\cal{S}}_2^{ip}$ and
 ${\cal{S}}_1^{op} \supset {\cal{S}}_2^{op}$.
 \hfill$\blacksquare$
\end{example}

\noindent
\textcolor{black}{The binary relation $``\succ"$ defined over a set of services is {\em not} an equivalence relation,
since, $``\succ"$ is not symmetric, i.e., if ${\cal{S}}_1 \succ {\cal{S}}_2$, then ${\cal{S}}_2 \nsucc {\cal{S}}_1$
(unless ${\cal{S}}_1 \simeq {\cal{S}}_2$).}
The dominance relationship further reduces the number of services.

Consider a service {{${\cal{S}}^{(1)}_i \in S^{(1)}$}} such that no other service in $S^{(1)}$ dominates ${\cal{S}}^{(1)}_i$.
${\cal{S}}^{(1)}_i$ then forms a group.
Consider {{$\{{\cal{S}}^{(1)}_{i_1}, {\cal{S}}^{(1)}_{i_2}, \ldots, {\cal{S}}^{(1)}_{i_k}\} \subset S^{(1)}$}},
such that {{${\cal{S}}^{(1)}_i \succ \{{\cal{S}}^{(1)}_{i_1}, {\cal{S}}^{(1)}_{i_2}, \ldots, {\cal{S}}^{(1)}_{i_k}\}$}}.
In this case, {{${\cal{S}}^{(1)}_{i_1}, {\cal{S}}^{(1)}_{i_2}, \ldots, {\cal{S}}^{(1)}_{i_k}$}} belong to the 
group formed by ${\cal{S}}^{(1)}_i$. The group is finally abstracted and represented by a single service ${\cal{S}}^{(2)}_i$. 
The inputs and outputs of ${\cal{S}}^{(2)}_i$ are same as in ${\cal{S}}^{(1)}_i$ respectively. The values of the QoS parameters 
of the dominating service are assigned to the values of the QoS parameters of 
${\cal{S}}^{(2)}_i$. However, no dominated service, in this abstraction step, can form a group. It may be noted that we always get a dominance relationship 
in the strict sense, since we have already combined the equivalent services in the previous abstraction level. 
With this abstraction step, we can still preserve the functional dependencies, while reducing the search space in the 
dependency graph. 


After the second level of abstraction, we have {{$S^{(2)} = \{{\cal{S}}^{(2)}_1, {\cal{S}}^{(2)}_2, \ldots, {\cal{S}}^{(2)}_{n_2}\}$}}. Each ${\cal{S}}^{(2)}_i \in S^{(2)}$ is an abstraction of a set of services of the first level, 
where one service dominates the rest. The set of services corresponding to ${\cal{S}}^{(2)}_i$ 
and ${\cal{S}}^{(2)}_j$ $(i \neq j)$ are not always mutually exclusive, 
since one service can be dominated by multiple services.
However, the number of services after the second level of abstraction still reduces. Therefore, the following lemma holds.

\begin{lemma}
 The number of services after the second level of abstraction is less than or equal to the total number of services 
 after the first level of abstraction, i.e., $n_2 \le n_1$. \hfill$\blacksquare$
\end{lemma}



\begin{example}
 Consider the example discussed in Section \ref{sec:overview}. As can be seen from the table in Figure \ref{fig:defInSR}, 
 the abstract service corresponding to $FS^{(1)}$ dominates the abstract service corresponding to $FS^{(2)}$. 
 Therefore, according to our abstraction mechanism, these two services are abstracted by a new abstract service, 
 say ${\cal{S}}^{(2)}_1$ 
 and we assign the values of the QoS parameters of the abstract service corresponding to $FS^{(1)}$ to ${\cal{S}}^{(2)}_1$.
 The number of services in the service repository reduces from 26 to 19.
 \hfill $\blacksquare$
\end{example}

\noindent
As earlier, we now can use any off-the-shelf composition method followed by a QoS constraint satisfying solution construction step. 
Since, in this level the QoS values of the dominant service are assigned to the abstract service, we cannot provide any 
guarantee regarding the optimality of the QoS parameters. However, this abstraction strategy also satisfies the preservation, soundness 
and no-loss properties by a transitive reasoning from the 
abstract service set at this level to the abstraction at the previous level. 

\subsubsection{Second level refinement}
\noindent
In this case as well, no solution may be found in the 
abstract space satisfying the QoS constraints, while a solution exists in the un-abstracted space before this abstraction 
step is executed. This necessitates refinement. 

\begin{example}
 Consider 6 services with two QoS parameters response time and throughput: 
 ${\cal{S}}^{(1)}_1: (100, 50)$, ${\cal{S}}^{(1)}_2: (75, 75)$, ${\cal{S}}^{(1)}_3: (150, 100)$, 
 ${\cal{S}}^{(1)}_4: (200, 150)$, ${\cal{S}}^{(1)}_5: (300, 50)$, ${\cal{S}}^{(1)}_6: (50, 150)$, such that
 ${\cal{S}}^{(1)}_1 \succ {\cal{S}}^{(1)}_2$, ${\cal{S}}^{(1)}_3 \succ {\cal{S}}^{(1)}_4$, 
 ${\cal{S}}^{(1)}_5 \succ {\cal{S}}^{(1)}_6$. According to dominance based abstraction, 
 ${\cal{S}}^{(1)}_1$ and ${\cal{S}}^{(1)}_2$ are abstracted by ${\cal{S}}^{(2)}_1$ with QoS values (100, 50);
 ${\cal{S}}^{(1)}_3$ and ${\cal{S}}^{(1)}_4$ are abstracted by ${\cal{S}}^{(2)}_2$ with QoS values (150, 100);
 ${\cal{S}}^{(1)}_5$ and ${\cal{S}}^{(1)}_6$ are abstracted by ${\cal{S}}^{(2)}_3$ with QoS values (300, 50).
 Also consider a query $\cal{Q}$ {\textcolor{black}{for which we either}} need to compose ${\cal{S}}^{(2)}_1$ and ${\cal{S}}^{(2)}_2$
 or ${\cal{S}}^{(2)}_1$ and ${\cal{S}}^{(2)}_3$. Now consider the QoS constraint as (150, 50). However, the solution,  
 returned by the {\textcolor{black}{optimal}} algorithm, has QoS values (250, 50) violating the QoS constraint.
 \hfill$\blacksquare$
\end{example}

\noindent
The QoS-based refinement here is similar to the previous level. The only difference, which we discuss below, lies in the selection step
presented in Line \ref{line:selection} of Algorithm \ref{algo:sub_first_refine}.
Consider {{$\{{\cal{S}}^{(1)}_i, {\cal{S}}^{(1)}_{i_1}, {\cal{S}}^{(1)}_{i_2}, \ldots, {\cal{S}}^{(1)}_{i_k}\}$}} be the services 
generated after the first abstraction step. Also consider, 
{{${\cal{S}}^{(1)}_i \prec \{{\cal{S}}^{(1)}_{i_1}, {\cal{S}}^{(1)}_{i_2}, \ldots, {\cal{S}}^{(1)}_{i_k}\}$}}.
In this abstraction phase, a new service ${\cal{S}}^{(2)}_i$ abstracts all these services, i.e., 
{{$\{{\cal{S}}^{(1)}_i, {\cal{S}}^{(1)}_{i_1}, {\cal{S}}^{(1)}_{i_2}, \ldots, {\cal{S}}^{(1)}_{i_k}\}$}} are abstracted by ${\cal{S}}^{(2)}_i$.
The input, output set of ${\cal{S}}^{(2)}_i$ are identical to the input, output set of ${\cal{S}}^{(1)}_i$. Therefore, it may be noted, some of the 
services corresponding to ${\cal{S}}^{(2)}_i$ may have more inputs than ${\cal{S}}^{(2)}_i$. 
While constructing a composite solution, if ${\cal{S}}^{(2)}_i$ is chosen in the solution, it must be based on its input set,
which implies at least one ${\cal{S}}^{(1)}_i$ corresponding to ${\cal{S}}^{(2)}_i$ can be activated by the set of inputs. It may be the case that
the other services, i.e., ${\cal{S}}^{(1)}_{ij}$, for {{$j = 1, 2, \ldots, k$}}, corresponding to ${\cal{S}}^{(2)}_i$ can also be 
activated based on the available set of inputs. However, the QoS values of ${\cal{S}}^{(1)}_i$ are assigned to the QoS of ${\cal{S}}^{(2)}_i$.
Therefore, while revising the values of the QoS parameters of ${\cal{S}}^{(2)}_i$, we need to look at all the services whose input sets are 
available in the system. In other words, while composing the set of services, we can only select a service whose inputs are available. The QoS based solution refinement technique deals with the $1^{st}$ level of abstract services 
corresponding to each $2^{nd}$ level abstract service of the solution. Therefore, after the QoS based solution refinement, 
we end up having the solution constructed over the $1^{st}$ level abstract services.

%
%
%
In this case as well, complete refinement may be needed to generate a solution, since the QoS based refinement step has the same limitation as earlier. 

\subsection{Abstraction of input equivalent services}
\noindent
We now proceed with the third level of abstraction based on input equivalence as defined in Definition \ref{def:inputEqui}.
Input equivalence is another functional characteristic which further reduces the search space.
The second level abstract services are divided into equivalence classes based on the binary equivalence relation $\simeq_i$. This forms a partition of $S^{(2)}$. Each equivalence class is then abstracted by a new service. 
Consider {{${\hat{S}}^{(2)}_i = \{{\cal{S}}^{(2)}_{i1}, {\cal{S}}^{(2)}_{i2}, \ldots, {\cal{S}}^{(2)}_{ik}\} \subset S^{(2)}$}} 
be an equivalence class, i.e., {{${\cal{S}}^{(2)}_{i1} \simeq_i {\cal{S}}^{(2)}_{i2} \simeq_i \ldots \simeq_i {\cal{S}}^{(2)}_{ik}$}}.
These are abstracted by ${\cal{S}}^{(3)}_{i}$. The input set of ${\cal{S}}^{(3)}_{i}$ is identical to the 
input set of any {{${\cal{S}}^{(2)}_{ij} \in {\hat{S}}^{(2)}_i$}}. The output set of ${\cal{S}}^{(3)}_{i}$ is the union of the output set of 
{{${\cal{S}}^{(2)}_{ij} \in {\hat{S}}^{(2)}_i$}}, i.e., {{${\cal{S}}^{(3)op}_{i} = \cup_{i = 1}^{k} ({\cal{S}}^{(2)op}_{ij})$}}, 
where {{${\cal{S}}^{(3)op}_{i}$}} represents the set of outputs of service ${\cal{S}}^{(3)}_{i}$. It may be noted that each service 
{{${\cal{S}}^{(2)}_{ij} \in {\hat{S}}^{(2)}_i$}} produces at least one unique output, i.e., this output is not produced by any service 
{\textcolor{black}{except}} {{${\cal{S}}^{(2)}_{ij} \in {\hat{S}}^{(2)}_i$}}. 
This is because we have already abstracted the services based on equivalence and dominance. 
Therefore, there does not exist any service pair which produces identical
set of outputs. Also there does not exist two services, such that the set of outputs of one is a superset of the same of another. 
Since the set of input-equivalent services can be activated simultaneously by the same set of inputs, they can be executed in parallel as well.
The QoS parameters of the abstract service are set as follows:

\begin{itemize}
 \item The response time as the maximum response time among all the services corresponding to it.
 \item The reliability (availability) value as product of the reliability (availability) of the corresponding services.
 \item The invocation cost as the sum of the invocation costs of all services corresponding to it.
\end{itemize}

\noindent
After the third level of abstraction, we have $S^{(3)} = \{{\cal{S}}^{(3)}_1, {\cal{S}}^{(3)}_2, \ldots, {\cal{S}}^{(3)}_{n_3}\}$.
Each {{${\cal{S}}^{(3)}_i \in S^{(3)}$}} is an abstract service corresponding to a set of input equivalent abstract services.
As in the first level of abstraction, in this case as well, 
the set of services corresponding to ${\cal{S}}^{(3)}_i$ and ${\cal{S}}^{(3)}_j$ are mutually exclusive for $i \neq j$ and the following lemma holds.

\begin{lemma}
 The number of services $n_3$ after the third level of abstraction is less than or equal to the number of services 
after the second level of abstraction, i.e., $n_3 \le n_2$. \hfill$\blacksquare$
\end{lemma}
%

\begin{example}
 Consider the example in Section \ref{sec:overview}. 
 As shown in the table of Figure \ref{fig:defInSR}, the abstract services corresponding to RS, SS, WF and 
 EM are input equivalent and thereby, according to our abstraction mechanism, 
 these services are abstracted by a new service, say ${\cal{S}}^{(3)}_1$ and we 
 assign the values of QoS parameters as discussed above.
 After this step, the number of services in the service repository reduces from 19 to 16.
\hfill $\blacksquare$
\end{example}

%
\noindent
As in the {\textcolor{black}{earlier}} abstractions, we can {\textcolor{black}{use}} any standard composition method on this abstract set.
Since the QoS values of the second layer abstract services in an equivalence class are considered to assign the QoS 
values {\textcolor{black}{to the}} corresponding abstract service (without considering whether each abstract service is required at runtime to serve a query), 
we cannot provide any guarantee on the optimal QoS values of the composite solution. 
However, in this case as well, we have a corresponding preservation, soundness, no-loss theorem.
%

\subsubsection{Third level of refinement}
\noindent
As in the previous levels, if we are able to construct a QoS constraint satisfying solution in the abstract space, we are done. However, if not, this abstraction step also necessitates a refinement. The QoS based solution 
refinement technique is similar in spirit to the ones discussed for the first two levels. Once we have a solution, we start 
traversing the solution backward, i.e., we start from the query outputs ${\cal{Q}}^{op}$ and 
find the services which produce these outputs. In this manner, we traverse till we get back the query inputs ${\cal{Q}}^{ip}$.
Once we encounter a service corresponding to more than one service in the previous level, we expand the service and check if 
there exists any {\em{unused services}}. As discussed, input-equivalent services produce at least one unique 
output. Sometimes, all these outputs may not be required to solve the query. Therefore, with respect to a query, there may exist
some unused services, whose unique outputs are not required to solve the query. We identify such services and modify the QoS 
parameters of the abstract service created in this level.
In other words, we recompute the values of the QoS parameters of the abstract service without considering the unused services
from the set of input equivalent services.
After QoS based solution refinement, 
we end up having the solution on the $2^{nd}$ level abstract services. In a similar note as above, a complete refinement may be required.
The complete refinement is similar as earlier.

\subsection{Abstraction using fusion}

\noindent
This is the final abstraction step. The abstraction is done by accumulating multiple functionally dependent services. The number of 
services generated in this level remain same as the number of services after the third level of abstraction. However, 
the dependency graph for a query is much smaller, as we discuss later in this section. Instead of composing 
multiple services at runtime, if a single abstract service can solve our purpose, that can directly impact the solution 
computation efficiency. 
%
In this level, we recursively consider all the services which can be activated by the outputs of an abstract service. Consider 
a service {{${\cal{S}}^{(3)}_{i} \in S^{(3)}$}} with {{$I^{(3)}_{i}$}} and {{$O^{(3)}_{i}$}} 
as the set of inputs and outputs respectively. In this abstraction, {{${\cal{S}}^{(3)}_{i}$}}
and the services functionally dependent on {{${\cal{S}}^{(3)}_{i}$}} are abstracted by a new service
{{${\cal{S}}^{(4)}_{i}$}}. 

Algorithms \ref{algo2} and \ref{algo1} show the construction of an abstract service 
{{${\cal{S}}^{(4)}_{i}$}}. The input set of {{${\cal{S}}^{(4)}_{i}$}} is same as the input set of 
{{${\cal{S}}^{(3)}_{i}$}} and the output set of {{${\cal{S}}^{(4)}_{i}$}} is the union of the set of 
outputs produced by the set of services corresponding to {{${\cal{S}}^{(4)}_{i}$}}.
For each abstract service in $S^{(3)}$, Algorithm \ref{algo2} is internally called from Algorithm \ref{algo1}.
{\textcolor{black}{The algorithms internally maintain a hashmap (Abstract), that stores the third level abstract services and their corresponding
dependent services as a key-value pair, as it traverses and encounters new service nodes in the dependency graph.}
Algorithm \ref{algo2} is a recursive algorithm which computes the dependency graph corresponding to a third layer 
abstract service. The worst case complexity of Algorithm \ref{algo1} is $O(n^2)$, where n is the number of services in the service repository. 
The worst case arises
when the previous abstraction level fails to reduce the number of abstract services.

For each abstract service ${\cal{S}}^{(3)}_{i} \in S^{(3)}$, we have an abstract service ${\cal{S}}^{(4)}_{i}$
corresponding to a set of services generated from ${\cal{S}}^{(3)}_{i}$. Intuitively, we construct a dependency graph starting 
from a service ${\cal{S}}^{(3)}_{i} \in S^{(3)}$. Therefore, each abstract service {{${\cal{S}}^{(4)}_{i} \in S^{(4)}$}} corresponds 
to a unique dependency graph constructed over the set of abstract services in $S^{(3)}$. In this case, it may be noted that the set of 
abstract services corresponding to ${\cal{S}}^{(4)}_{i}$ and ${\cal{S}}^{(4)}_{j}$ are not mutually exclusive. 
We compute the QoS parameters of the dependency graph corresponding to ${\cal{S}}^{(4)}_{j}$
and assign these values to ${\cal{S}}^{(4)}_{j}$.

\begin{algorithm}
\scriptsize
  \caption{RecursiveConstruction}
  \begin{algorithmic}[1]
   \State {Input: Service ${\cal{S}}^{(3)}_i$}
   \State $A = \{{\cal{S}}^{(3)}_i\}$; $I^{*} = I^{(3)}_i$;
   \For {each service ${\cal{S}}^{(3)}_j \in (S^{(3)} \setminus A)$ and is activated by $I^{*}$} 
      \If {Abstract[${\cal{S}}^{(3)}_i$] == null}
	\State RecursiveConstruction (${\cal{S}}^{(3)}_j$);
      \EndIf
      \State $A \leftarrow A \cup Abstract[{\cal{S}}^{(3)}_i]$;
      \For {$s \in A$}
	  \State $I^{*}$ = $I^{*} \cup$ Output set of $s$;
      \EndFor
   \EndFor 
   \State $A$ is abstracted by ${\cal{S}}^{(4)}_i$;
   \State $I^{(4)}_i = I^{(3)}_i$;
   \For {$s \in A$}
      \State $O^{(4)}_i = O^{(4)}_i \cup$ Output set of $s$;
   \EndFor
   \State $Abstract[{\cal{S}}^{(3)}_i] = A$;
  \end{algorithmic}
  \label{algo2}
\end{algorithm}
\vspace{-0.6cm}
\begin{algorithm}
\scriptsize
  \caption{AbstractServiceConstruction}
  \begin{algorithmic}[1]
   \State Initialize Abstract[${\cal{S}}^{(3)}_i$] = null, $\forall {\cal{S}}^{(3)}_i \in S^{(3)}$;
   \For {each ${\cal{S}}^{(3)}_i \in S^{(3)}$ if Abstract[${\cal{S}}^{(3)}_i$] $\ne$ null}
      \State RecursiveConstruction (${\cal{S}}^{(3)}_i$);
   \EndFor   
  \end{algorithmic}
  \label{algo1}
\end{algorithm}

%
%
%

\begin{figure}[!htb]
 \centering
 \includegraphics[width=\linewidth]{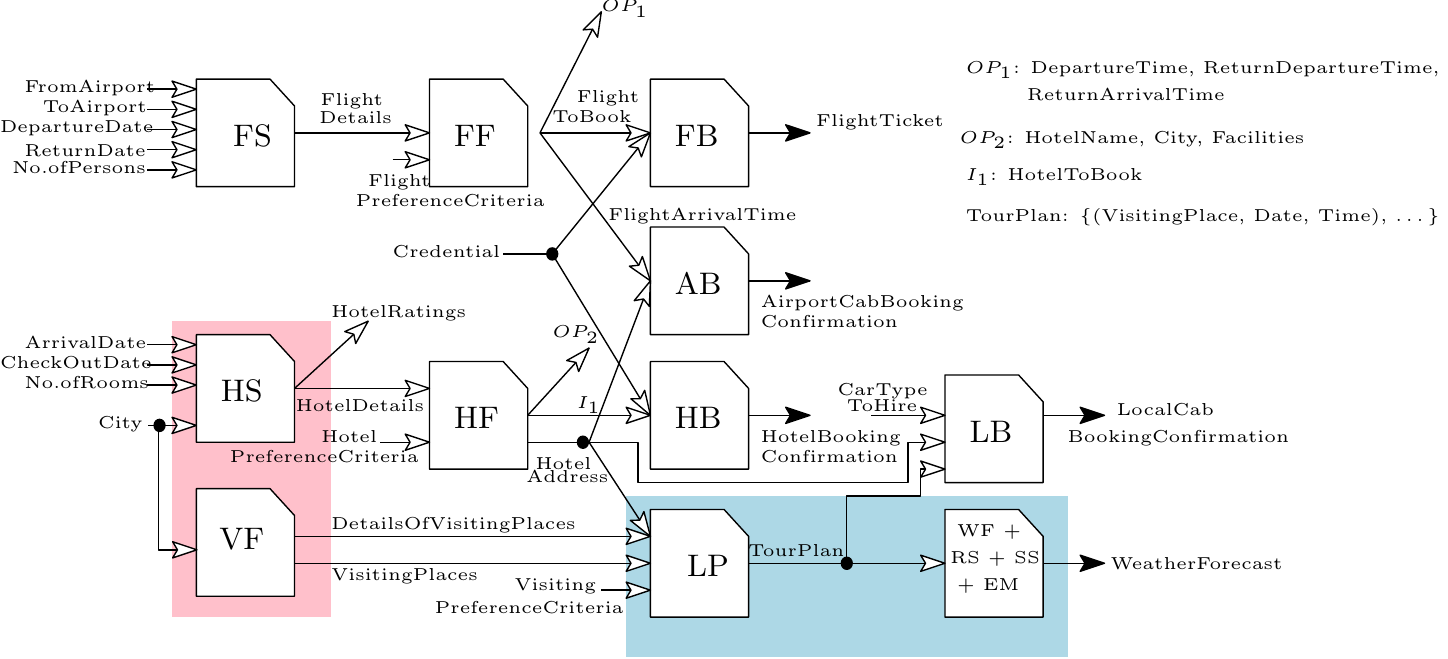}
 \caption{Abstract service generation using fusion}
 \label{fig:depdSR}
 \end{figure}
 
 \begin{example}
 Consider the example in Section \ref{sec:overview}. 
 The construction of an abstract service based on fusion is shown in Figure \ref{fig:depdSR}.
 The colored {\textcolor{black}{rectangular boxes}} show different dependency graphs corresponding to each abstract service created at this level.
 We start with each service in the previous level and compute the services which are functionally dependent on that service.
 We assign the values of the QoS parameters of the overall dependency graph to the corresponding abstract service as discussed above.
 The number of services to be considered for composition now reduces from 16 to 14 after this abstraction step is executed.
 \hfill$\blacksquare$
\end{example}


%

\subsubsection{Composition with abstract services {\textcolor{black}{after fusion}}}
\noindent
For constructing the composition on the abstract service set obtained after fusion, we use the notion of a sub-service, which we formally define below. 

\begin{definition}{\em [\bf Sub-service: ]}
 ${\cal{S}}^{(4)}_{i}$ is a sub-service of ${\cal{S}}^{(4)}_{j}$, if the set of services corresponding to
 ${\cal{S}}^{(4)}_{i}$ is a subset of the set of services corresponding to ${\cal{S}}^{(4)}_{j}$.
 \hfill$\blacksquare$
\end{definition}

\noindent
Once the query comes into the system, the dependency graph is constructed. 
While constructing the dependency graph using the abstract services in $S^{(4)}$,
 if a service $s \in S^{(4)}$ is used to construct the graph, no sub-service of $s$ is used for the 
 dependency graph construction. Though this abstraction does not reduce the number of abstract
 services in this level, the size of the dependency graph reduces due to elimination of the sub services. 
  Similar to the third level of abstraction, here also we cannot provide any QoS optimality guarantee on the composite solution.
 The preservation, soundness and the no-loss theorems {\textcolor{black}{still continue to hold}}.

\begin{figure}[!htb]
\centering
\includegraphics[width=\linewidth]{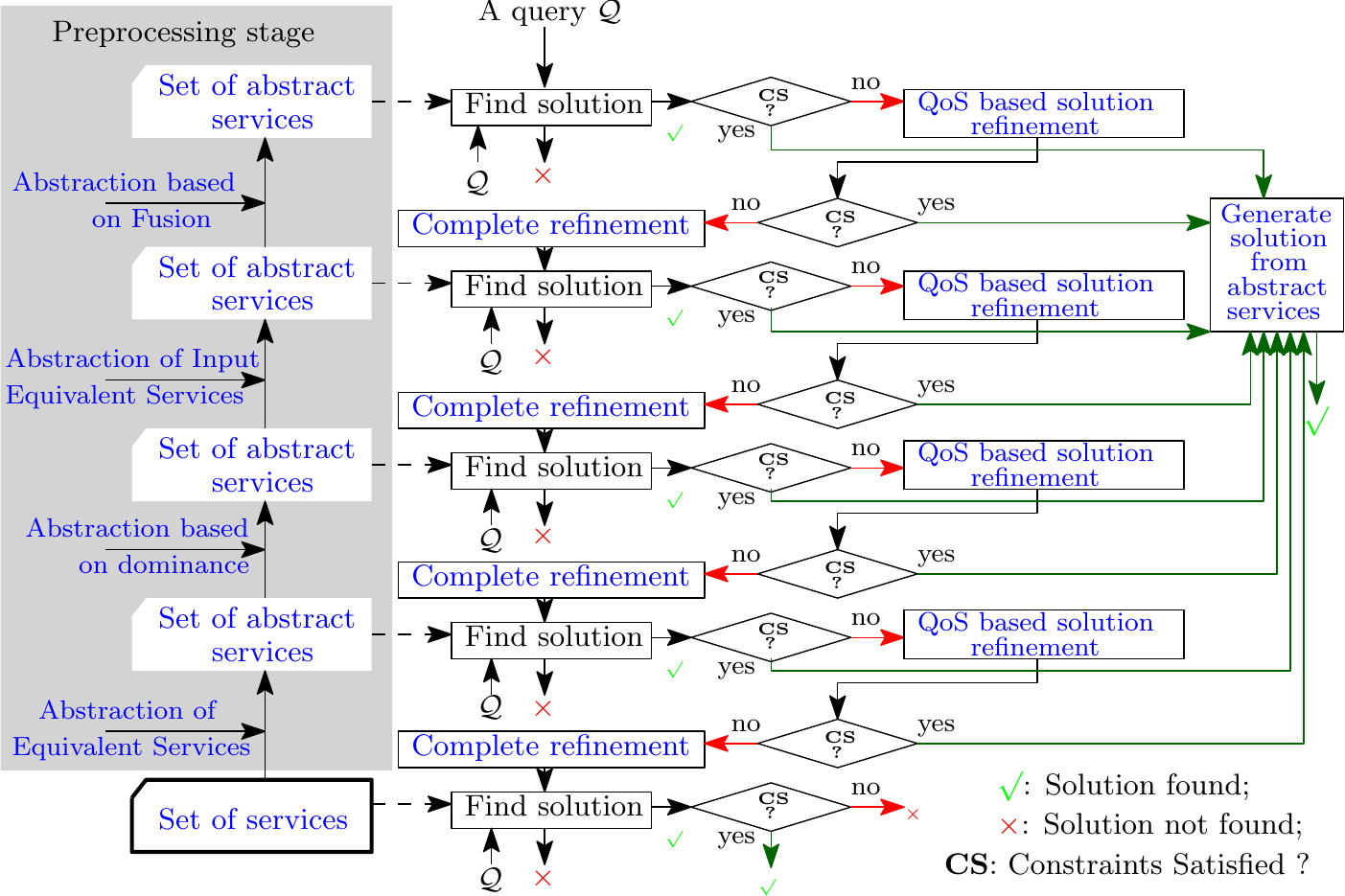}
\caption{Flow of composition}
\label{fig:flow}
\end{figure}
 
\subsubsection{The Final refinement step}
\noindent
We first discuss QoS based solution refinement followed by the complete refinement.
{\textcolor{black}{This involves the following:}}

\begin{itemize}
 \item We start traversing the solution in the forward direction. We start with the query inputs and find the services that are activated 
	by the query inputs. We then find the services that are activated by the query input as well as the outputs of the services that
	we have already considered. In this way we proceed until we get back the set of query outputs.
 \item Once we encounter an abstract service ${\cal{S}}^{(4)}_i$ whose corresponding dependency graph consists of more than one service,
	we traverse the dependency graph and remove all the services which are {\textcolor{black}{redundant to solve the given query, i.e., without
	which the query can still be answered.}} 
	We then compute the values of the QoS parameters and assign these to ${\cal{S}}^{(4)}_i$.
 \item We recompute the QoS parameters of the solution.
\end{itemize}
\noindent
The QoS based solution refinement technique above deals with the $4^{th}$ level abstract services 
corresponding to each $3^{rd}$ level abstract service of the solution. Therefore, after the QoS based solution refinement {\textcolor{black}{step}}, 
we end up having the solution constructed over the $3^{rd}$ level abstract services.
In the complete refinement strategy, we replace the abstract services by the services in the 
third level of abstraction in the dependency graph and compute the solution.

\subsection{The overall flow}
\noindent

\noindent
Figure \ref{fig:flow} shows the complete flow of our architecture. 
The {\em{Find solution}} block generates {\textcolor{red}{\textbf{$\times$}}}, when no solutions are found that 
satisfy the functional dependency and generates {\textcolor{green}{\textbf{$\surd$}}}, if at least one solution 
is found which satisfies the functional dependency. 
At any level, once we have the abstract set of services, we can use any standard 
technique for service composition. Once we get a solution in terms of the abstract services, we need to return the solution in terms of the original services. This can be done by replacing the abstract service in the solution at any level with the original services it represents at the previous level, continuing to the first level. Algorithm~\ref{algo:revert} shows the the solution reconstruction.

 \begin{algorithm}
 \scriptsize
   \caption{SolutionConstruction}
   \begin{algorithmic}[1]
    \State {Input: Solution in terms of abstract services (Sol)}
    \State {Output: Solution in terms of original services}
    \For {each service $s \in Sol$}      
   \Comment{$s \in S^{(4)}$}
       \State Replace $s$ by the corresponding dependency graph;
    \EndFor
    \For {each service $s \in Sol$}
    \Comment{$s \in S^{(3)}$}
       \State Replace by the set of services corresponding to $s$;
    \EndFor
    \For {each service $s \in Sol$}
    \Comment{$s \in S^{(2)}$}
       \State Replace by the dominant service corresponding to $s$;
    \EndFor
    \For {each service $s \in Sol$}
    \Comment{$s \in S^{(1)}$}
       \State Replace by the service corresponding to $s$ whose QoS parameters are \indent assigned to $s$;
    \EndFor
   \end{algorithmic}
  \label{algo:revert}
\end{algorithm}

%

\section{Experimental Results}\label{sec:result}
Our proposed algorithms were implemented\footnote{Available at: http://www.isical.ac.in/$\sim$soumi\_r/scResearch.html} 
in Java (version 1.7.0\_60, 32 bit). All experiments were performed on 
a 2.53GHz machine with 4GB DDR3 RAM. 

\subsection{Data sets}
\noindent
We evaluated our methods on the following data sets:
\begin{itemize}
 \item 5 public repositories of the WSC-2009 \cite{bansal2009wsc} dataset
 \item 19 repositories of the ICEBE-2005 WSC \cite{icebe2005} dataset
 \item An extended version of the description in Section \ref{sec:overview}.
\end{itemize}

\noindent
{\textcolor{black}{To demonstrate the power of abstraction and the strength of our methods, we implemented our proposal on top of i) a single QoS aware optimal composition algorithm~\cite{xia2013web}, ii) a multiple QoS aware optimal composition algorithm~\cite{DBLPChattopadhyayB16} and iii) a heuristic approach~\cite{6009375}. }}
In the following subsections, we show the performance gain enabled by our method over all the cases. In our experiments, we considered the following 
problems for a given query:

\begin{itemize}
 \item Generating all feasible solutions
 \item Generating a solution with a single QoS parameter
 \item Generating a solution with multiple parameters
\end{itemize}



\begin{figure*}[!htb]
 \centering
 (a)\includegraphics[height=3.5cm,width=5cm]{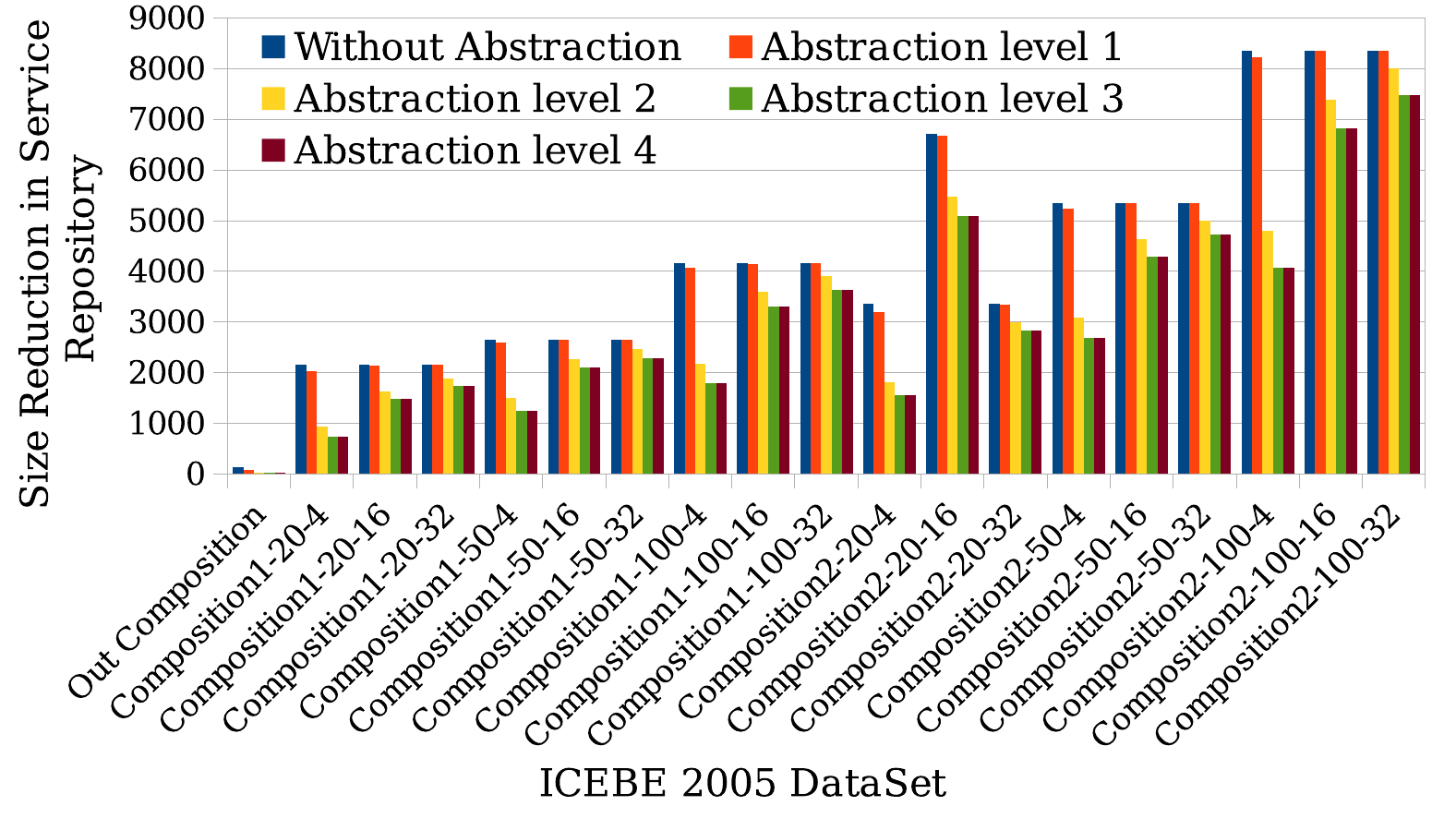}
 (b)\includegraphics[height=3.5cm,width=5cm]{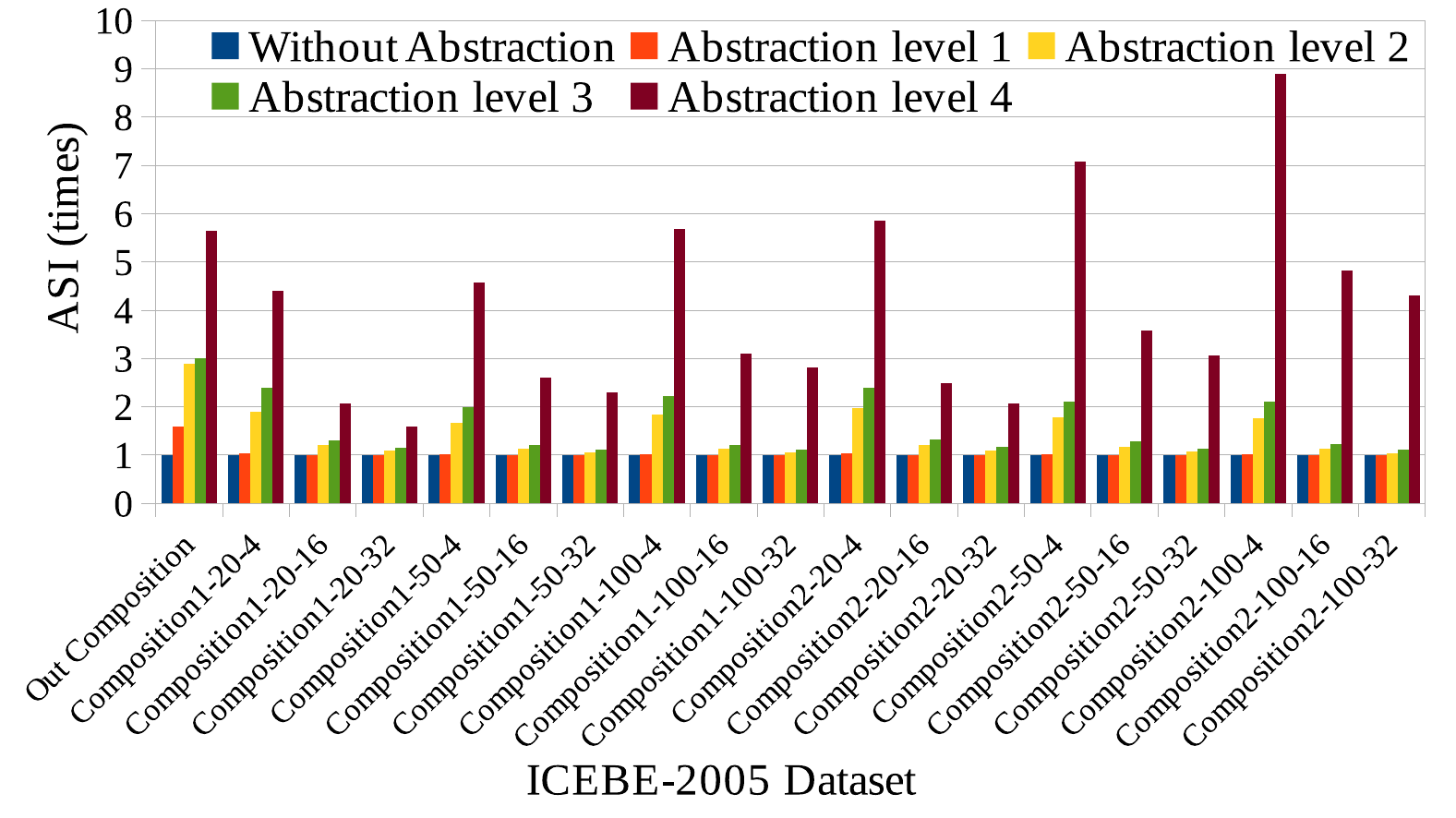}
 (c)\includegraphics[height=3.5cm,width=5cm]{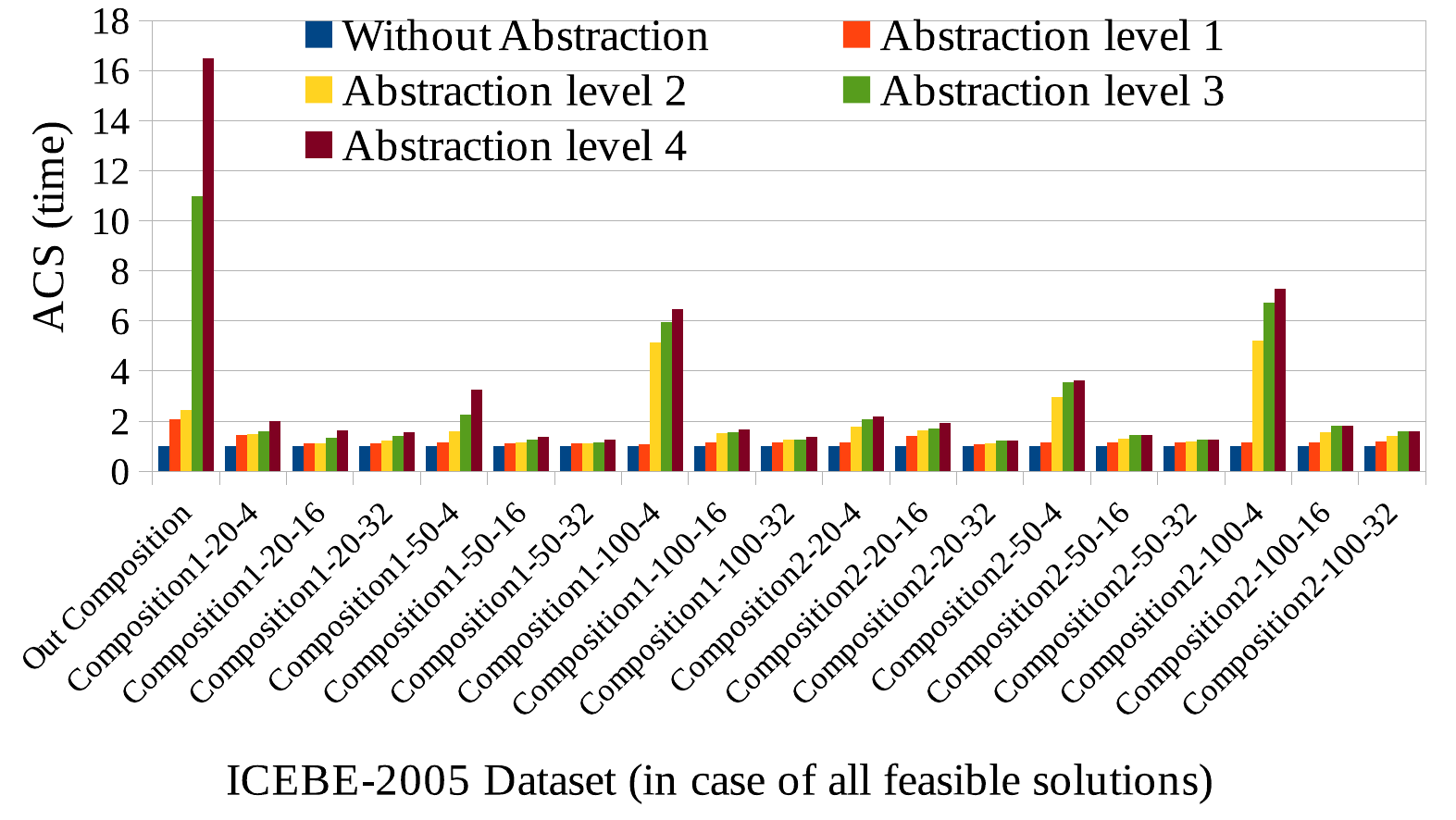}
 \caption{\scriptsize{ICEBE dataset: (a)No. of services reduction in repository (b) {\textcolor{black}{ASI}} in dependency graph 
 (c) {\textcolor{black}{ACS}} to construct dependency graph}}
\label{fig:SRSizeICEBE}
\end{figure*}
\begin{figure*}[!htb]
 \centering
 (a)\includegraphics[height=3.5cm,width=5cm]{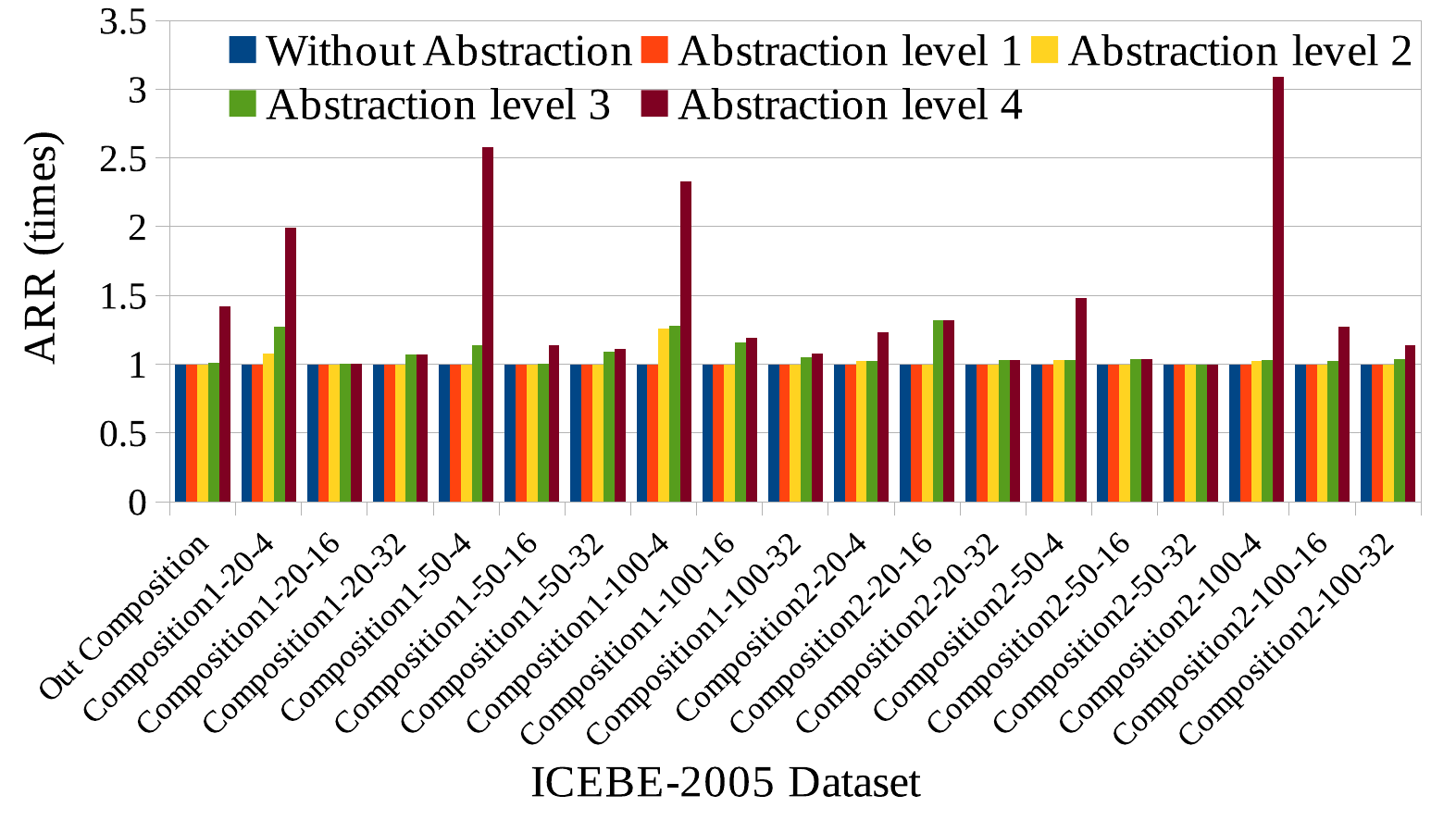}
 (b)\includegraphics[height=3.5cm,width=5cm]{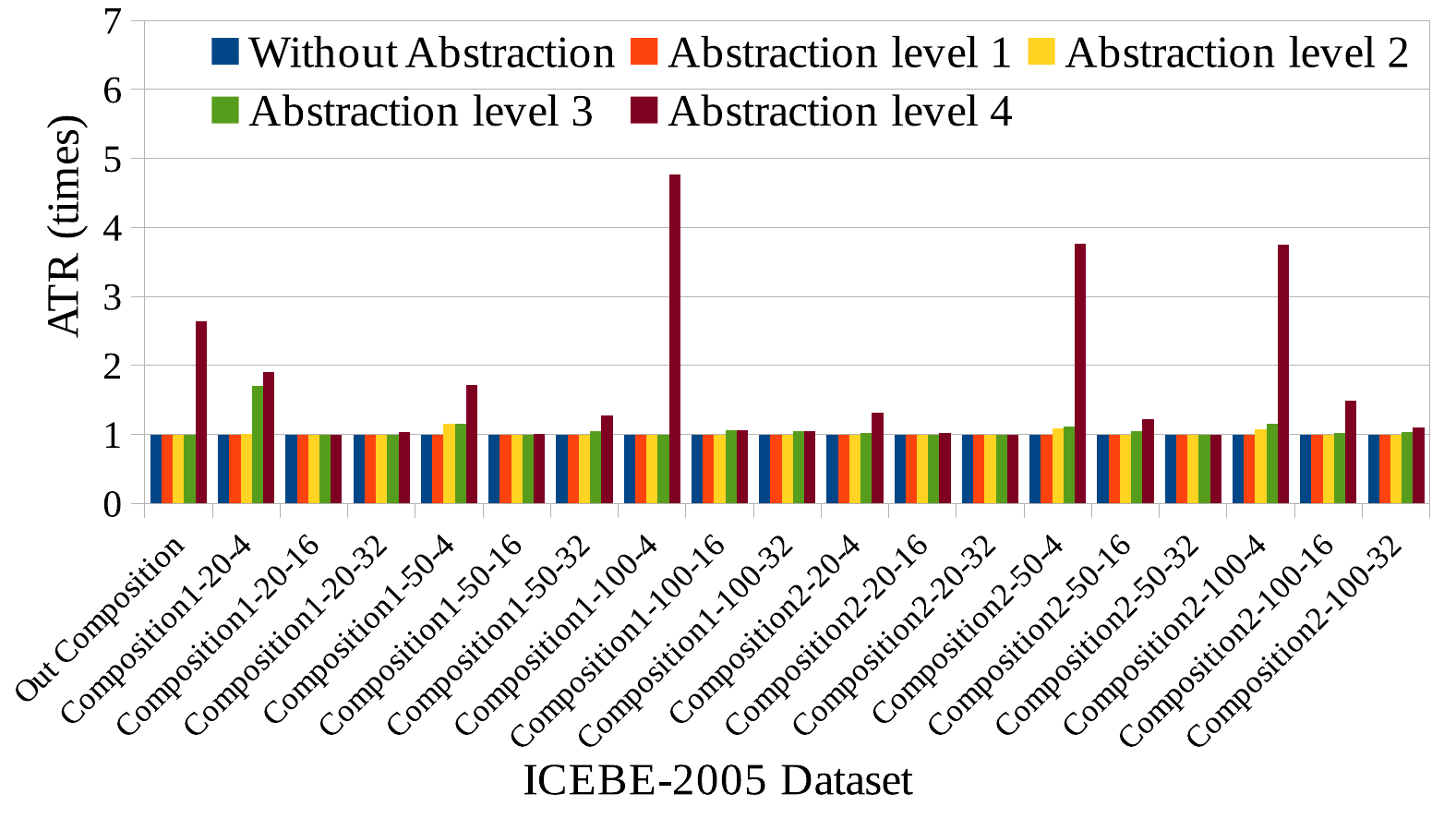}
 (c)\includegraphics[height=3.5cm,width=5cm]{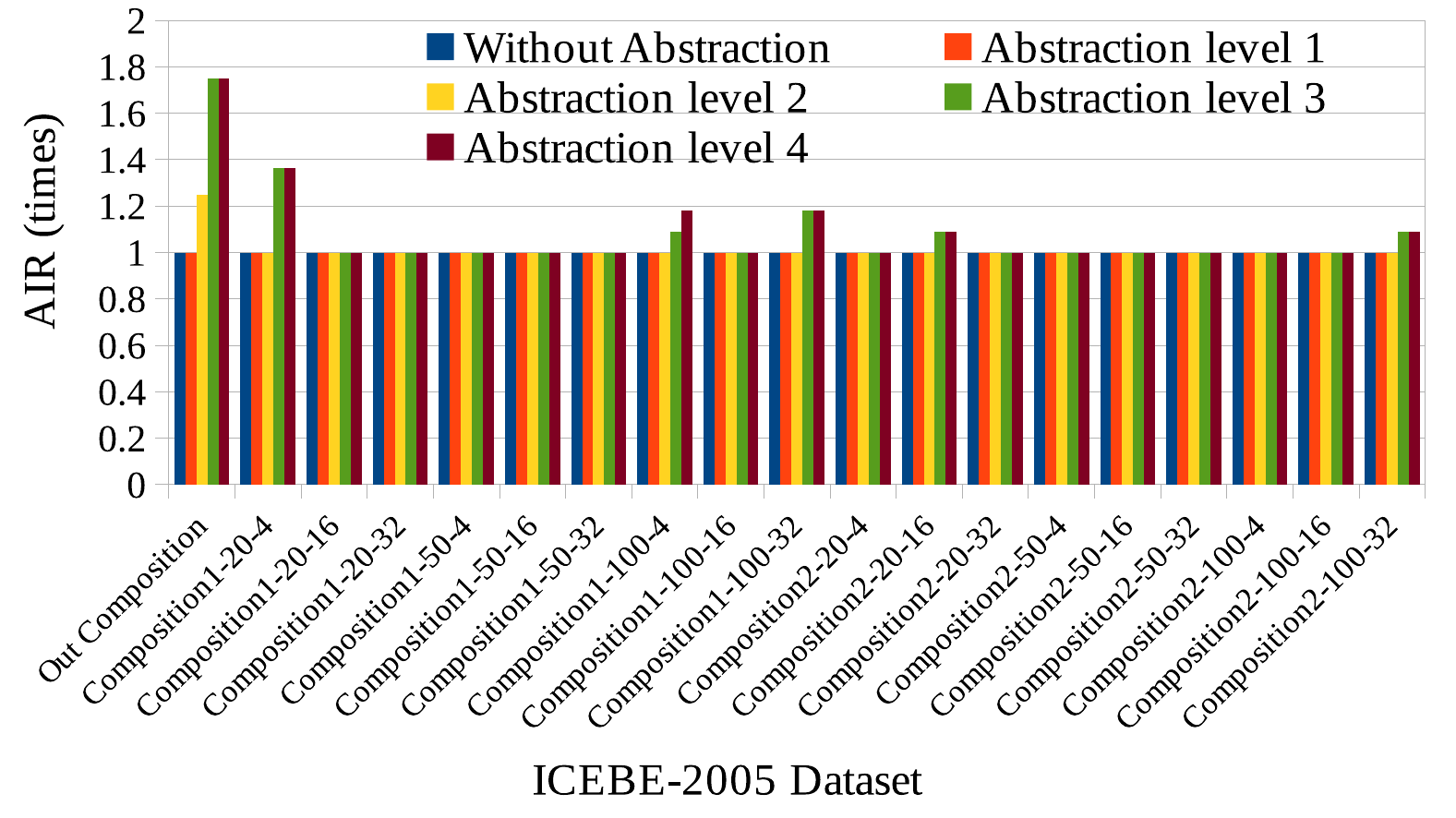}
 \caption{\scriptsize{ICEBE dataset (a) ARR (b) ATR (c) AIR}}
\label{fig:RTICEBE}
\end{figure*}

\begin{figure*}[!htb]
 \centering
 (a)\includegraphics[height=3.5cm,width=5cm]{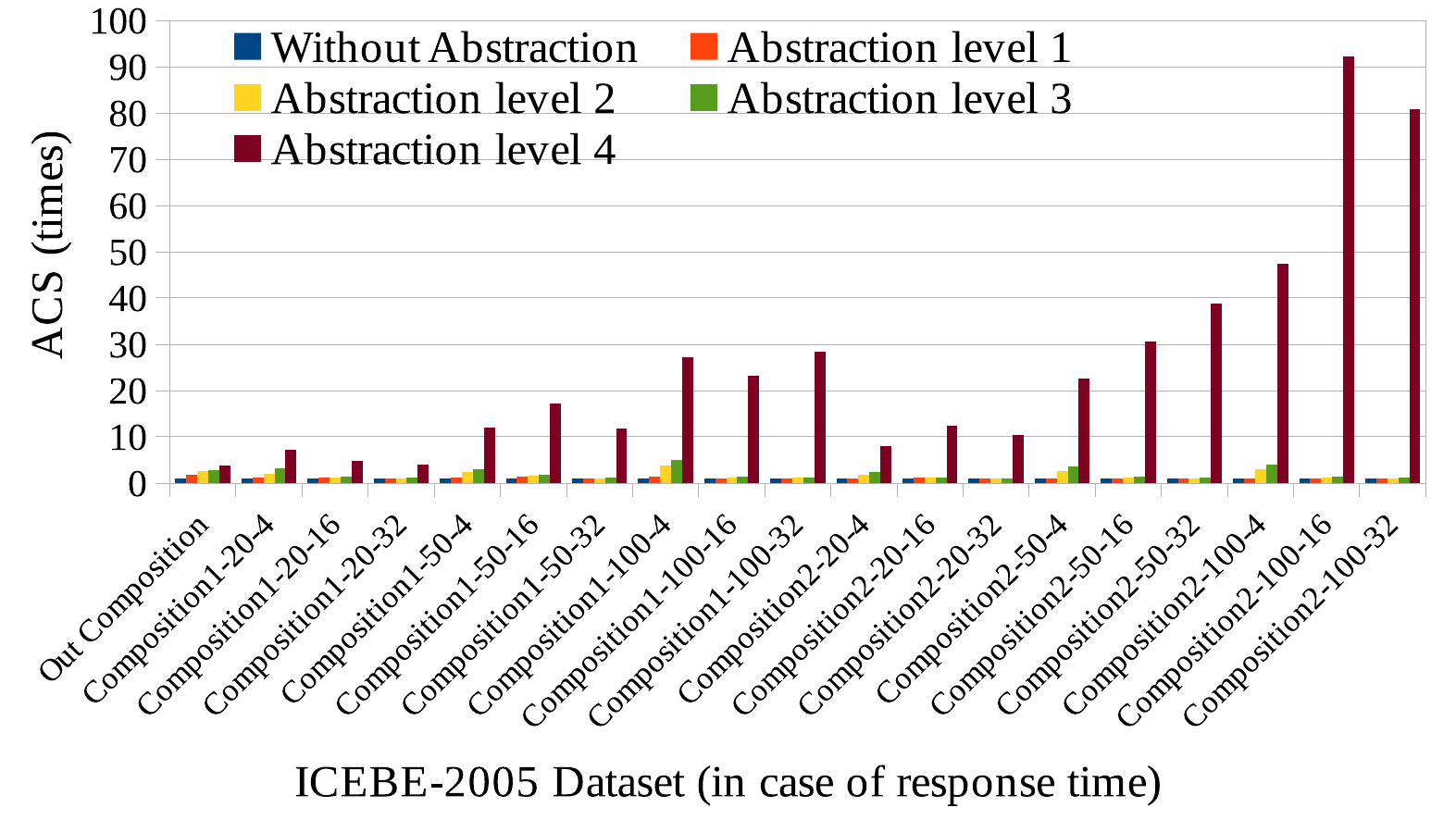}
 (b)\includegraphics[height=3.5cm,width=5cm]{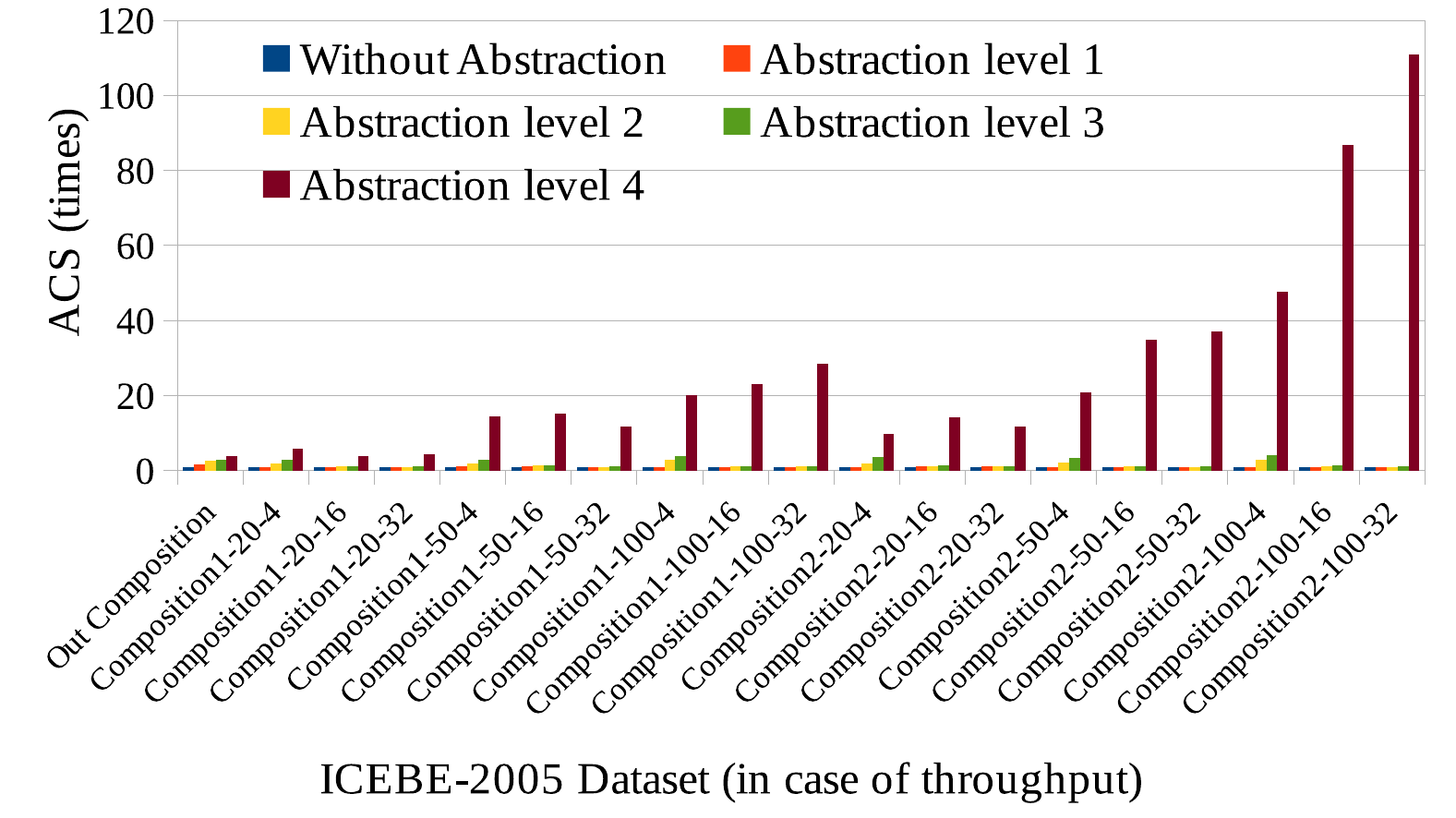}
 (c)\includegraphics[height=3.5cm,width=5cm]{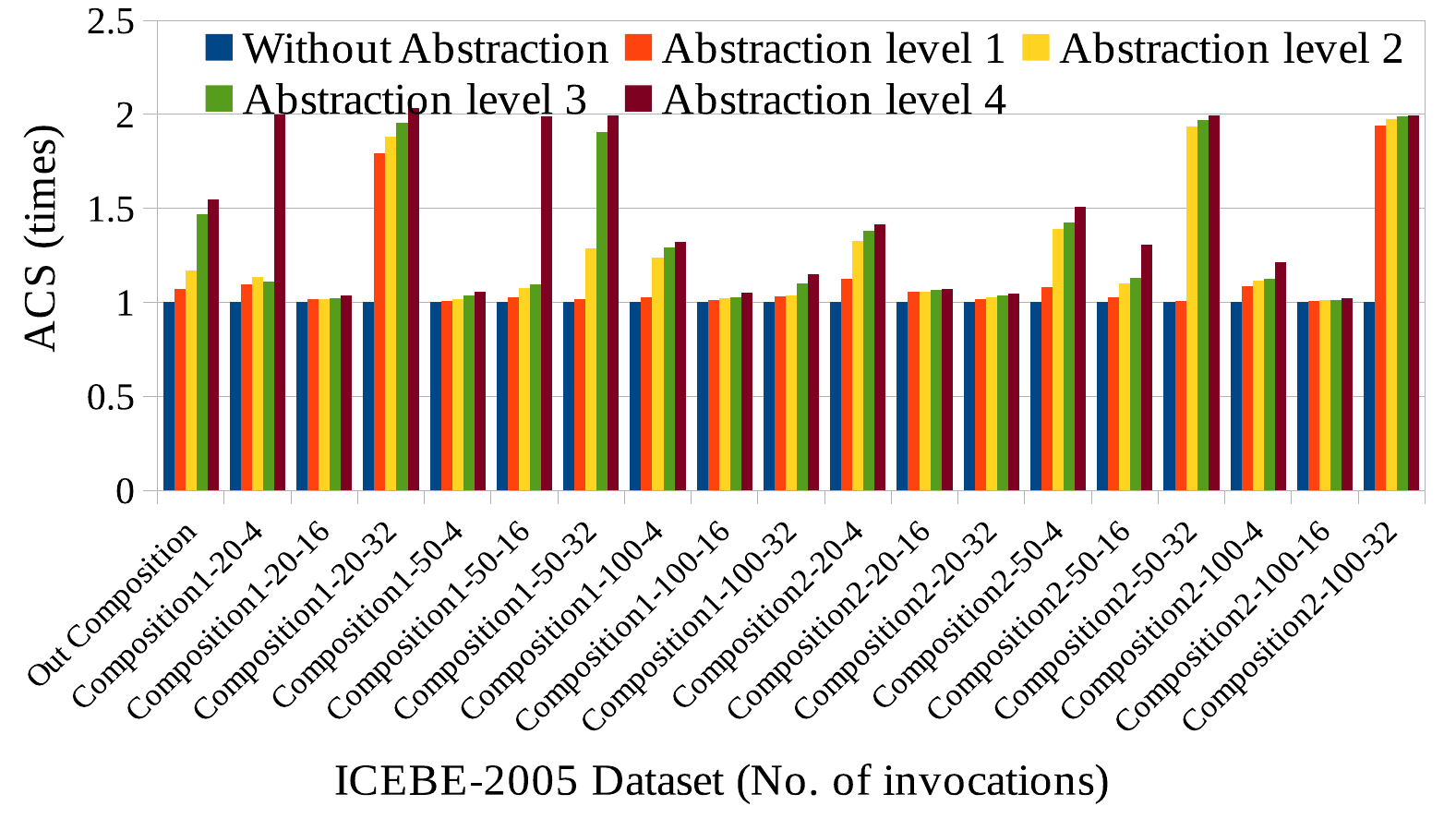}
 \caption{\scriptsize{ICEBE dataset average speedup to construct the solution (a) response time aware (b) throughput aware (c) no. of invocations aware}}
\label{fig:RTTimeICEBE}
\end{figure*}

\subsection{Evaluation}
\noindent
We first discuss the performance metrics considered for comparison between the approaches.


\noindent{\scriptsize
$\text{Average response time ratio (ARR)} = AVG_{{\cal{Q}}} (\frac{\text{RT computed with abstraction}}{\text{RT computed without abstraction}})$\\

\noindent
Average throughput ratio (ATR) = $AVG_{{\cal{Q}}} (\frac{\text{TR computed without abstraction}}{\text{TR computed with abstraction}})$\\

\noindent
Average no. of invocations ratio (AIR) = $AVG_{{\cal{Q}}} (\frac{\text{NI computed with abstraction}}{\text{NI computed without abstraction}})$\\

\noindent
Average size improvement  (ASI) = $AVG_{{\cal{Q}}} (\frac{\text{NSD without abstraction}}{\text{NSD  with abstraction}})$\\

\noindent
Average computation speedup (ACS) = $AVG_{{\cal{Q}}} (\frac{\text{CT without abstraction}}{\text{CT with abstraction}})$ \\
}

\noindent
where, RT, TR, NI, CT, NSD and AVG stand for response time, throughput, the number of invocations, computation time, the number of services 
in the dependency graph and average respectively and $\cal{Q}$ represents the set of queries.

%
%
%

\begin{table}[!ht]
\tiny
\caption{\tiny{ICEBE-2005\_Composition1 dataset: with multiple QoS constraints}}
\centering
\begin{tabular}{|c|c|c|c|c|}
 \hline
   Data   & \#       & Abstraction & \# memory out & Avg. composition\\
   sets   & queries  & level       & errors        & time (ms)       \\
   \hline\hline   
   && 0 & 7 & 12195\\   \cline{3-5}
   20-4 &  11 &  1 & 4 & 3662.26\\   \cline{3-5}
   && 2 & 0 & 4214.36\\   \cline{3-5}
   && 3 & 0 & 2521.0 \\   \cline{3-5}
   && 4 & 0 & 59.64\\
   \hline\hline
   && 0, 1, 2, 3 & 11 & -\\   \cline{3-5}
   20-16 &  11 & 4 & 1 & 17095.3\\
   \hline\hline
   && 0, 1, 2, 3 & 11 & -\\   \cline{3-5}
   20-32 &  11 & 4 & 6 & 452.804\\
   \hline\hline
   && 0, 1 & 11 & -\\   \cline{3-5}
   50-4 &  11 & 2 & 7 & 13868.5\\   \cline{3-5}
   && 3 & 6 & 9076.804 \\   \cline{3-5}
   && 4 & 0 & 29.18\\
   \hline\hline
   && 0, 1, 2, 3 & 11 & -\\   \cline{3-5}
   50-16 &  11 & 4 & 1 & 3222.604\\
   \hline\hline
   && 0, 1. 2, 3 & 11 & -\\   \cline{3-5}
   50-32 &  11 & 4 & 3 & 1844.37\\
   \hline\hline
   && 0, 1 & 11 & -\\   \cline{3-5}
   100-4 &  11 & 2 & 10 & 115.94\\   \cline{3-5}
   && 3 & 10 & 232 \\   \cline{3-5}
   && 4 & 0 & 27.637\\
   \hline\hline
   && 0, 1, 2, 3, 4 & 11 & -\\   \cline{3-5}
   100-16 &  11 & 4 & 0 & 826.397\\
   \hline\hline
   && 0, 1, 2, 3 & 11 & -\\   \cline{3-5}
   100-32 &  11 & 4 & 1 & 12015.99\\
   \hline\hline
\end{tabular}\label{tab:wmc_ILP_comp1}
\end{table}

\begin{table}[!ht]
\tiny
\caption{\tiny{ICEBE\_2005\_Composition2 dataset: with multiple QoS constraints}}
\centering
\begin{tabular}{|c|c|c|c|c|}
 \hline
   Data   & \#       & Abstraction & \# memory out & Avg. composition\\
   sets   & queries  & level       & errors        & time (ms)       \\
   \hline\hline
   && 0, 1 & 11 & -\\   \cline{3-5}
   20-4 &  11 & 2 & 10 & 67.496\\   \cline{3-5}
   && 3 & 10 & 7.095 \\   \cline{3-5}
   && 4 & 0 & 83.73\\
   \hline\hline
   && 0, 1, 2, 3 & 11 & -\\   \cline{3-5}
   20-16 &  11 & 4 & 1 & 1874.09\\
   \hline\hline
   && 0, 1, 2, 3 & 11 & -\\   \cline{3-5}
   20-32 &  11 & 4 & 4 & 1983.245\\
   \hline\hline
   && 0, 1 & 11 & -\\   \cline{3-5}
   50-4 &  11 &  2 & 10 & 27.5\\   \cline{3-5}
   && 3 & 10 & 3.894 \\   \cline{3-5}
   && 4 & 0 & 136.82\\
   \hline\hline
   && 0. 1, 2, 3 & 11 & -\\   \cline{3-5}
   50-16 &  11 & 4 & 2 & 747.36\\
   \hline\hline
   && 0, 1, 2, 3 & 11 & -\\   \cline{3-5}
   50-32 &  11 & 4 & 4 & 34928.85\\
   \hline\hline
   && 0, 1, 2, 3 & 11 & -\\   \cline{3-5}
   100-4 &  11 & 4 & 10 & 777.97\\   \cline{3-5}
   && 4 & 7 & 197.321\\
   \hline\hline
   && 0, 1, 2, 3 & 11 & -\\   \cline{3-5}
   100-16 &  11 & 4 & 6 & 2737.91\\
   \hline\hline
   && 0, 1, 2, 3 & 11 & -\\   \cline{3-5}
   100-32 &  11 & 5 & 4 & 234981.35\\
   \hline\hline
\end{tabular}\label{tab:wmc_ILP_comp2}
\end{table}

\noindent
We now discuss the evaluation result on 3 different datasets.

\subsubsection{Evaluation on ICEBE WSC-2005 Benchmark {\textcolor{black}{dataset}}}
\noindent
Figure \ref{fig:SRSizeICEBE}(a) shows the reduction in the number of services in the service repository across different abstraction levels for the 
ICEBE-2005 WSC dataset. This is achieved at preprocessing / design time, before arrival of a query.
As evident from the figure, the number of services reduces in each level except in the last level (as discussed in Section~\ref{sec:method}).




We first discuss the performance gain achieved by our method at runtime, when implemented on top of \cite{xia2013web}.
Figure \ref{fig:SRSizeICEBE}(b) shows the average reduction in the number of services in the dependency graph across 
different abstraction levels in response to a set of 11 queries that are provided as part of the ICEBE dataset. 
The X axis of the figure represents the dataset, while the Y axis represents ASI across different abstraction levels. 
As evident from the figure, there is a significant reduction in the number of services in each level, the average reduction across
different levels are 1.04, 1.39, 4.05, 4.13 times respectively. It is worth noting that though in the last abstraction level, 
the reduction in the number of services in the service repository is not visible, however, the reduction is quite significant 
(4.13 times on an average) in dependency graph construction in response 
to a query. Figure \ref{fig:SRSizeICEBE}(c) shows ACS achieved in dependency graph construction
across different abstraction levels, the average speedup gained across different levels are 1.21, 1.9, 2.64, 3.13 times
respectively.

\begin{figure*}[!htb]
 \centering
 (a) \includegraphics[height=3cm,width=5cm]{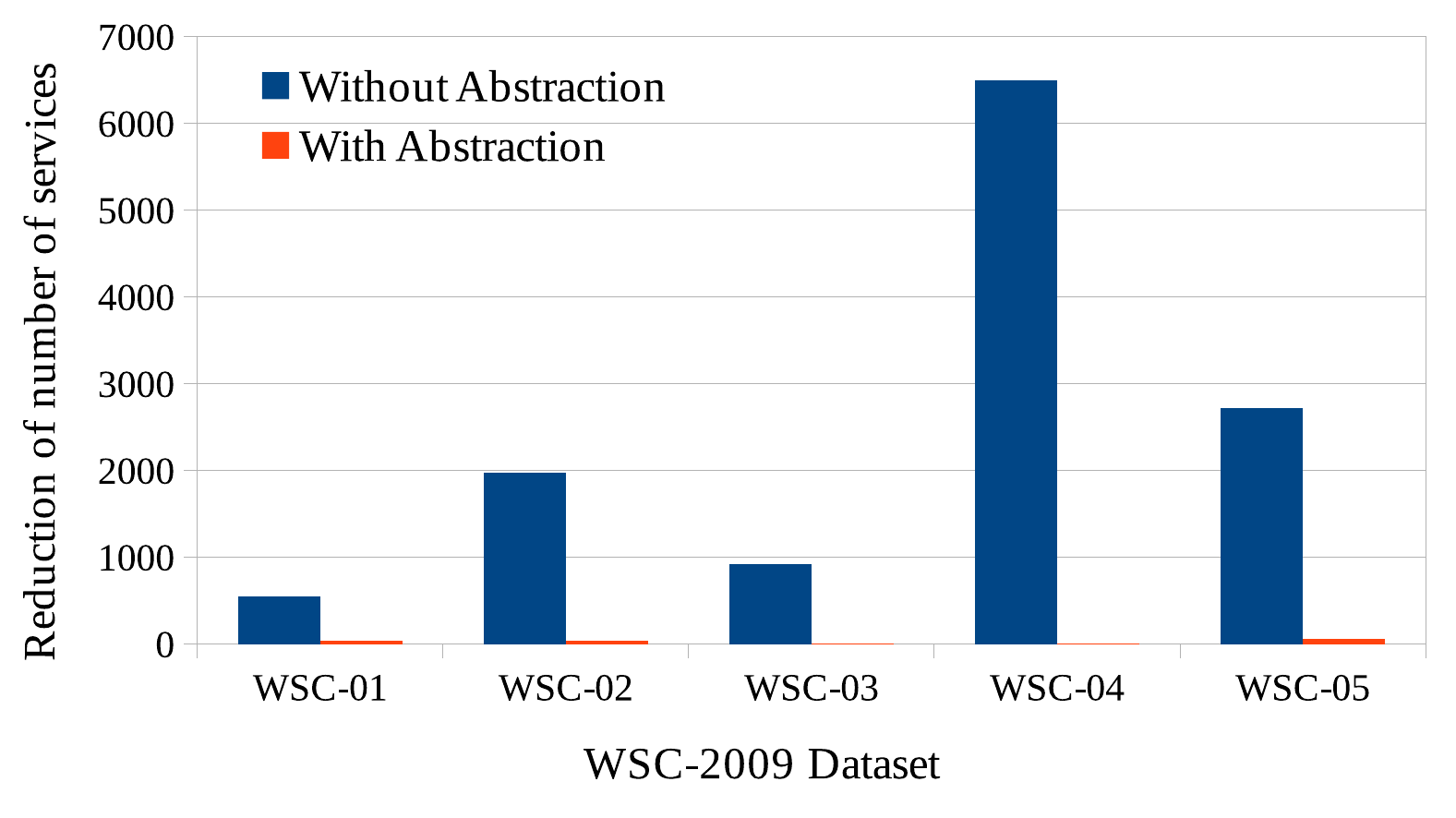}
 (b) \includegraphics[height=3cm,width=5cm]{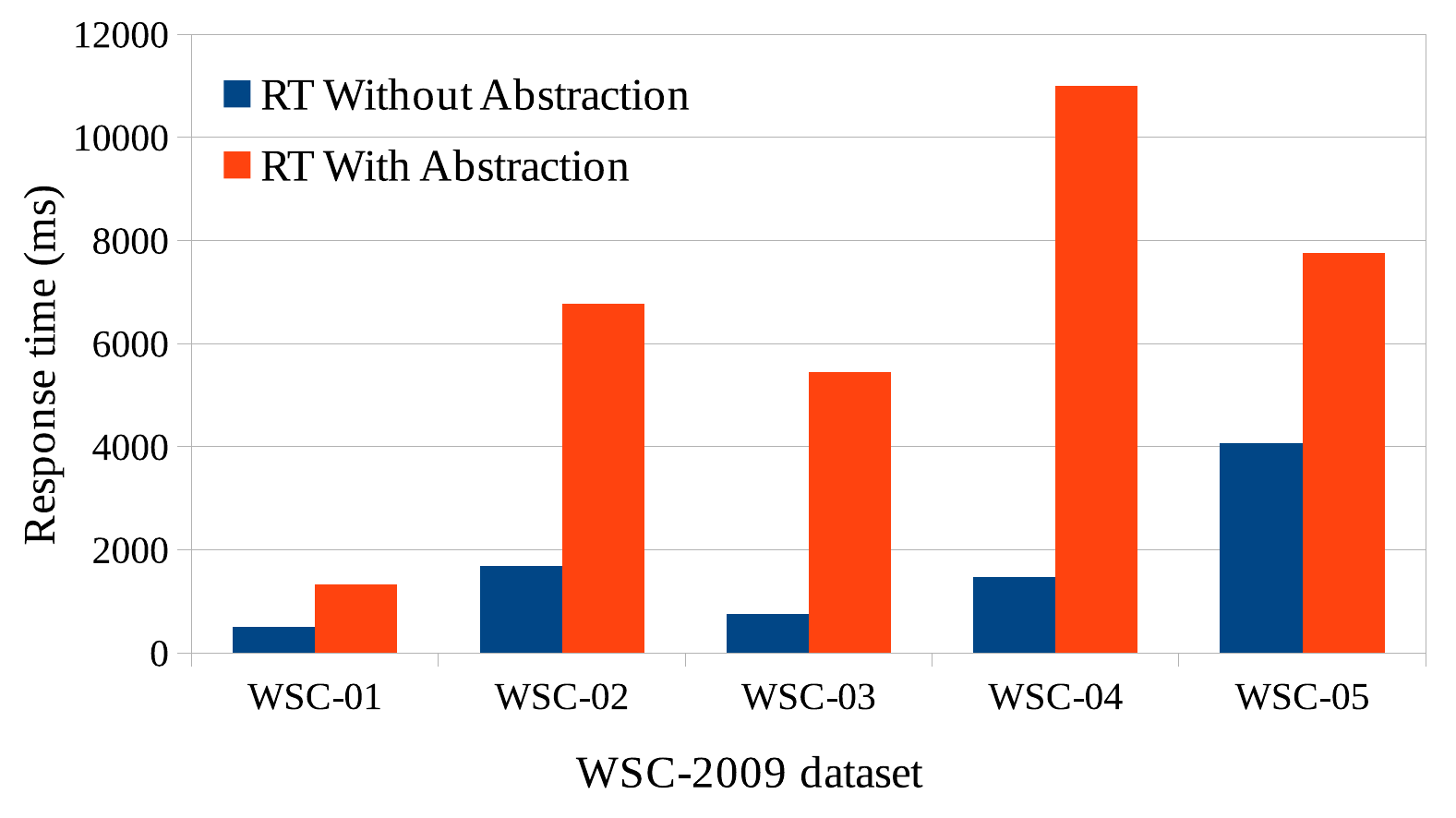}
 (c) \includegraphics[height=3cm,width=5cm]{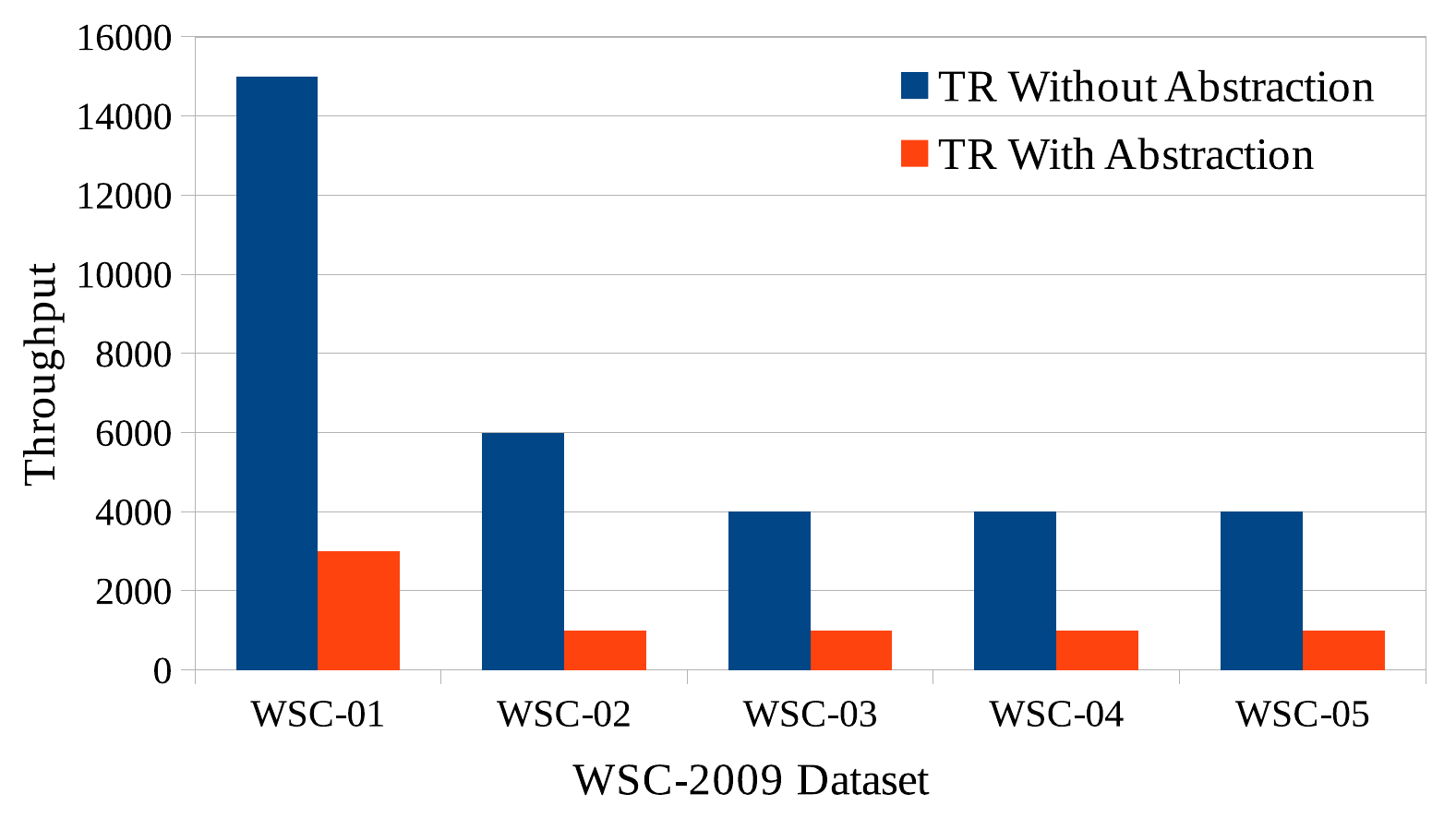}
 \caption{\scriptsize{WSC-2009 dataset: (a) {\textcolor{black}{Size reduction in dependency graph}} (b) Response time (c) Throughput}}
 \label{fig:wsc}
 \end{figure*}
 
 \begin{figure*}[!htb]
 \centering
 (a) \includegraphics[height=3cm,width=5cm]{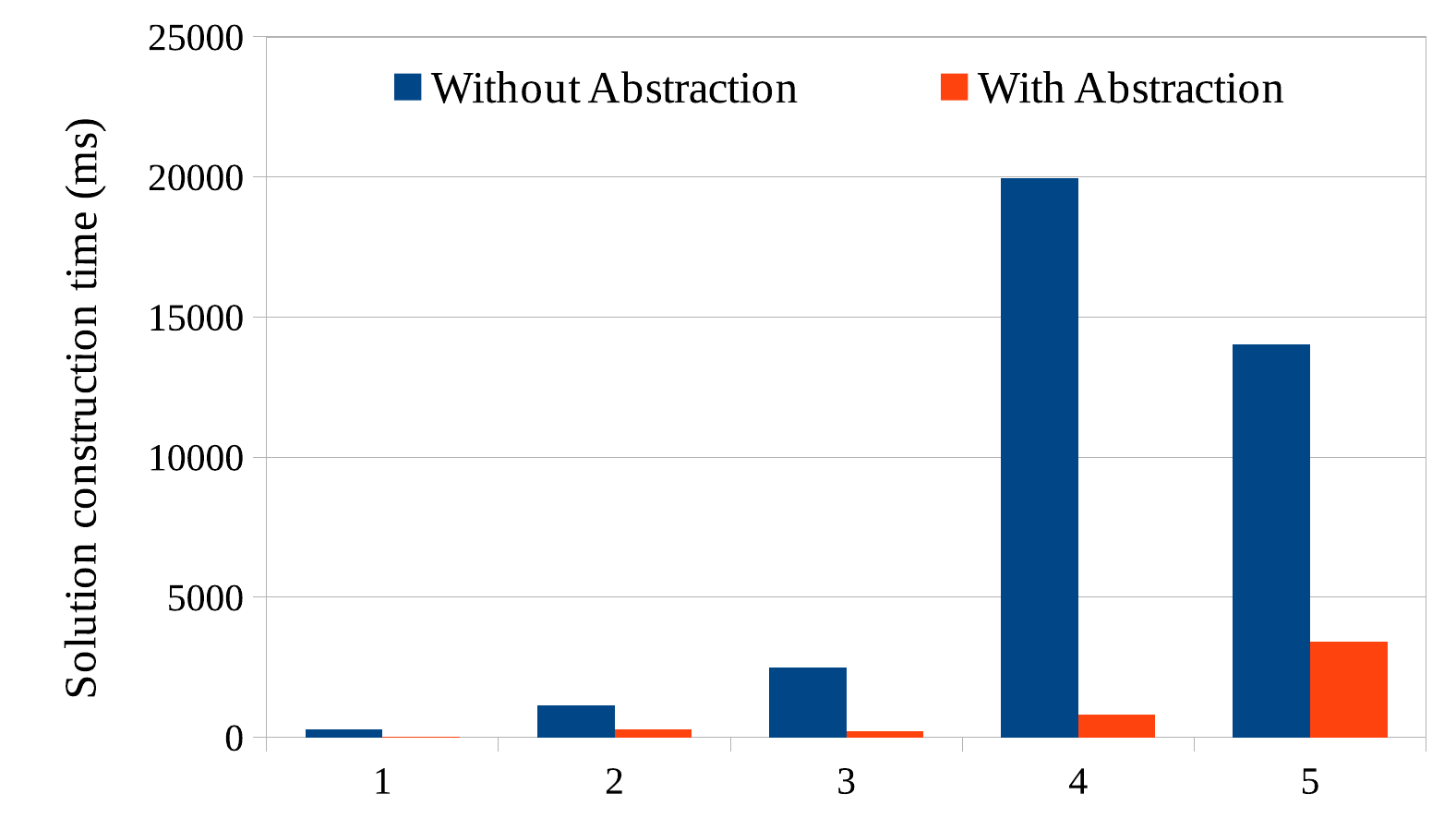}
 (b) \includegraphics[height=3cm,width=5cm]{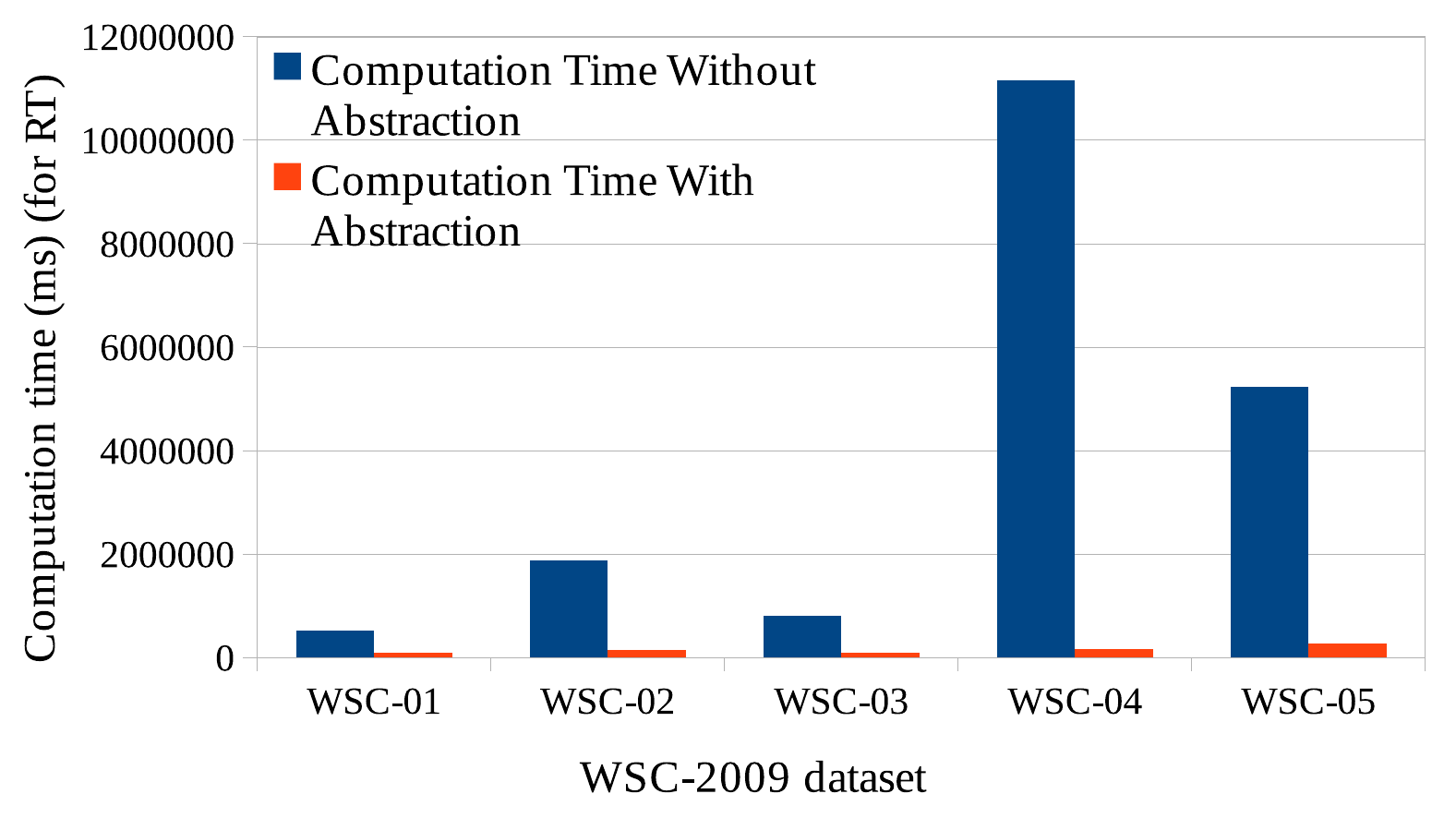}
 (c) \includegraphics[height=3cm,width=5cm]{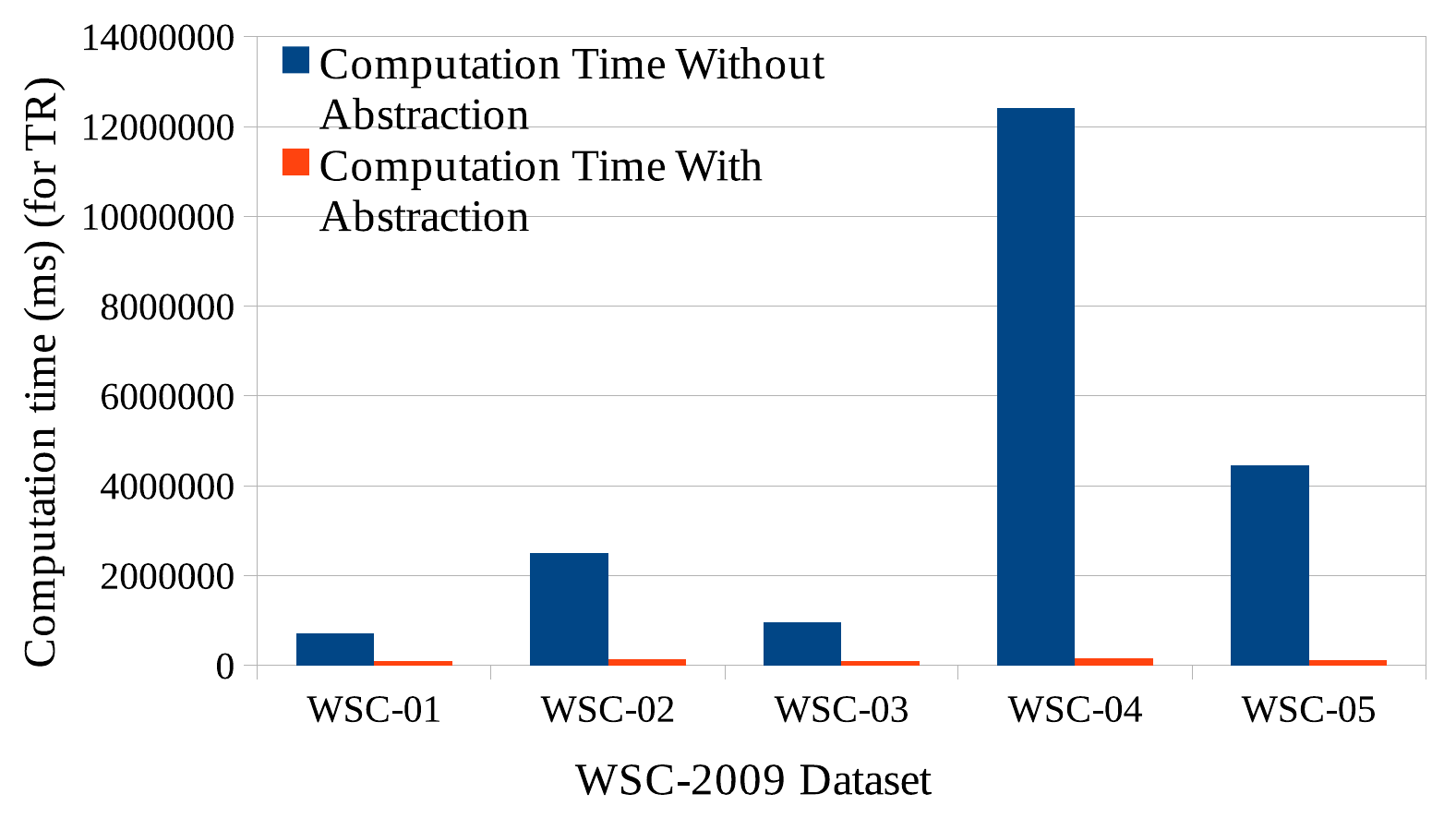}
 \caption{\scriptsize{WSC-2009 dataset: (a) Dependency graph construction time (b) Computation time for RT (c) Computation time for TR}}
 \label{fig:wscTime}
 \end{figure*}

%


{\textcolor{black}{Since}} our approach solves the composition problem on the abstract service groups, there is not always a guarantee 
of optimal solution generation, for the second abstraction level onwards, as discussed in Section~\ref{sec:method}. We show the 
degradation in QoS values empirically on standard benchmark datasets in presence of these abstractions. 
The ICEBE WSC-2005 dataset does not contain values of the QoS parameters of the services. Therefore, for each service we generated the values of the response time and throughput following a normal distribution. Figure \ref{fig:RTICEBE}(a) shows ARR across 
different abstraction levels in response to the 11 queries. Figure \ref{fig:RTTimeICEBE}(a) shows the ACS achieved in 
constructing the optimal response time with different abstraction levels on top of \cite{xia2013web}. Evidently, there is a significant 
speedup gain in computation time as compared to the degradation in solution quality. For 2 out of the 19 datasets, our method is able to 
derive the optimal response time even after all four levels of abstraction.

%

Figure \ref{fig:RTICEBE}(b) shows the ATR across different abstraction levels in response to the 11 queries. 
Figure \ref{fig:RTTimeICEBE}(b) shows the corresponding ACS with different abstraction levels on top of \cite{xia2013web}.
In this case as well, for 3 out of the 19 datasets, we still get the optimal throughput after all abstractions.

%


We now compare our proposal with a heuristic method \cite{6009375}, which considers the number of invocations as a QoS parameter.
Figure \ref{fig:RTICEBE}(c) shows a plot of AIR with different abstraction levels in response to the same set of 11 queries, while 
Figure \ref{fig:RTTimeICEBE}(c) shows the corresponding ACS achieved on top of \cite{6009375}.
Evidently, there is a significant speedup gain in computation time as compared to the change in the number of invocations. 
In this case, our methods generate the same solution (i.e., without degrading the average number of invocations) 
for 13 out of the 19 datasets after all abstractions.

%




\noindent
Finally, we consider multiple QoS parameters for the ICEBE-2005 datasets and the results are shown in Tables \ref{tab:wmc_ILP_comp1} and
\ref{tab:wmc_ILP_comp2}. We consider response time, reliability, availability, invocation cost 
and the number of service invocations as QoS parameters. The QoS values of each service are generated randomly.
Columns 2, 3, 4 and 5 of the table represent the number of queries, 
the abstraction level, the number of memory out errors for each abstraction and average time for composition respectively.
As evident from the tables, the average composition time decreases considerably as the abstraction level increases. 
It may be noted that in many cases,  
the ILP-based optimal algorithm \cite{DBLPChattopadhyayB16} gets memory out errors, as shown in Tables \ref{tab:wmc_ILP_comp1} 
and \ref{tab:wmc_ILP_comp2}.
For example, consider the 20-4 dataset of Table \ref{tab:wmc_ILP_comp1}. For the case without abstraction, the ILP fails to 
generate any solution (because of memory-out error) for 7 queries. However, as the abstraction level increases, the number 
of memory out error decreases.

 

\subsubsection{Evaluation on the WSC-2009 dataset}
\noindent
Figure \ref{fig:wsc}(a) shows the reduction achieved in the number of services for dependency graph construction with and without
abstraction in response to a query, while Figure \ref{fig:wscTime}(a) shows the corresponding computation speedup.
It is evident from the figures, in presence of abstraction, the solution generation time is considerably less. For the WSC-2009 dataset, 
we did not observe any service reduction for first 3 levels of abstraction and hence, omitted them.
 

WSC-2009 contains the values of response time and throughput of the services. Figure \ref{fig:wsc}(b) shows the response times with 
and without abstraction in response to a query in our method, while Figure \ref{fig:wscTime}(b) shows the computation time needed to 
construct the optimal response time with and without abstraction on top of \cite{xia2013web}.
As evident from the figures, there is a significant speedup gain in computation time as compared to the degradation in optimal response time. 
A similar gain is achieved for throughput. Figure \ref{fig:wsc}(c) shows the throughput with and without abstraction in response to a query 
and Figure \ref{fig:wscTime}(c) shows the corresponding computation time required to generate the optimal throughput 
with and without abstraction using \cite{xia2013web}.

%
 \noindent




\subsubsection{Evaluation on a synthetic dataset}
\noindent
We extended the service description discussed in Section \ref{sec:overview}. 
%
We considered 30 different service categories. Each service category performs a specific operation / task. Under each category, 
there are 3 or 4 different sub categories. Each sub category is selected based on 
input-output parameters. The services under a specific sub category have identical set of inputs and outputs.
We used an in-house web crawler and the open travel alliance \footnote{http://www.opentravel.org/} dataset to get the number of services 
for some service categories (e.g., searchFlight, bookFlight, searchHotel,
bookHotel, forecastWeather, bookAirportTransport, bookLocalTransport, searchRestaurant etc). For the remaining service
categories, we randomly generated the number of services. Once this was done,
we randomly divided the services into multiple sub categories. 
We considered the query as discussed in Section \ref{sec:overview}.
The total number of services in the service repository was 2461. We used the 
QWS\footnote{http://www.uoguelph.ca/ qmahmoud/qws/index.html/} dataset to assign the QoS values to the services. The QWS dataset has 8 different QoS parameters and more than 2500 services. From the QWS dataset, we randomly selected 2461
services and the corresponding QoS values were assigned to the services in our repository.
\begin{table}[htb]
 \tiny
 \caption{\scriptsize{{\textcolor{black}{Synthetic}} dataset: Performance of our approach}}
  \begin{center}
  \begin{tabular}{c|c|c}
  \hline
   SL & Speed up (times) & Abstraction Level \\
   \hline
   1 & 728 & 4\\\hline
   2 & 722 & 4\\\hline
   3 & 513 & 4\\ \hline
   4 & 426 & 3\\\hline
   5 & 196 & 2\\\hline
   6 & 122 & 1\\\hline
   7 & 69  & 1\\\hline
   8 & 1.23 & 1\\\hline
   9 & -6.45 & 0\\\hline
   10 & -13 & 0\\
   \hline
  \end{tabular}\label{tab:ourData}
 \end{center}
\end{table}
\noindent
The service repository initially contained 2461 services. We had 82, 30, 24 and 24 services after the first, 
second, third and fourth levels of abstraction respectively. The dependency graph with respect to the query consisted of 1423 services. 
We had 35, 12, 10 and 8 services after the first, second, third and fourth levels of abstraction respectively.
We first generated a random QoS constraint such that the fourth level abstract services can produce the solution satisfying 
all QoS constraints. We gradually tightened the constraints, till no solutions existed, as shown in the 10$^{th}$ row of Table \ref{tab:ourData}. Table \ref{tab:ourData} presents the performance 
of our approach. A larger row number indicates tighter constraints. Column 2 of Table \ref{tab:ourData} represents
the speed up with respect to the solution generated without any abstraction, Column 3 presents at which abstraction level,
we get a solution to the query satisfying all QoS constraints. As evident from the table, as abstraction level decreases, 
the speed up also decreases. In the final two rows (8$^{th}$ and 9$^{th}$), the solution is generated without abstraction, 
therefore we observe performance degradation. 



%
%
\section{Related Work}\label{related}
\noindent
{\textcolor{black}{
A significant amount of work has been done by considering different perspectives of service composition and discovery
\cite{7226855,jiang2012continuous,zhang2013selecting,mehdi2012trustworthy}.}}
The primary objective of service composition methods has been the computation 
of the optimal service composition result~\cite{rodriguez2015hybrid,schuller2012cost,xia2013web,chen2012redundant} considering functional 
and non-functional attributes. The optimality requirement however, in general, has proved to be an expensive 
requirement~\cite{lecue2009towards,chattopadhyay2015scalable} for service composition solutions. 
Therefore, heuristic solutions~\cite{wuqos,6009375,guidara2015heuristic,el2010tqos} have been proposed, 
that have the ability to generate solutions fast and handle large and complex service spaces~\cite{lecue2009towards}, but have sub-optimal solution quality~\cite{song2011workflow,pistore2005automated}. Table \ref{tab:relatedWork} presents a summary of some of the popular approaches in 
service composition based on different factors considered.

\begin{table*}[htb]
 \tiny
 \caption{\scriptsize{Description of State-of-the-Art Approaches}}
  \begin{center}
  \begin{tabular}{|c|c|c|c|c|c|c|c|}
   \hline
   References & Focus & Single /        & QoS Params & Static / & Optimal / & Deterministic / & Method \\
          &       & Multiple Params & Considered & Dynamic  & Heuristic & probabilistic   & \\
   \hline\hline
   \cite{lecue2009towards} & Scalability & Multiple & RT, EP & Static & QoS constrained & Deterministic & Constraint Satisfaction Problem\\
   \hline
   \cite{rodriguez2015hybrid} & Redundancy control  & Both & RT, TR & Static & Optimal & Deterministic & Graph based search\\
   \hline
   \cite{schuller2012cost} & Service Composition & Multiple & Cost, RT, TR, AV  & Static & Optimal + QoS constrained & Stochastic & ILP, Greedy Heuristic\\
                           &                     &          &                   &        & Heuristic                 &            & \\  
   \hline
   \cite{aiello2009optimal} & Service Composition & Single & RT, TR & Static & Optimal & Deterministic & Graph based search\\
   \hline
   \cite{6009375} & Service Composition & Single & NI & Static & Heuristic & Deterministic & A*\\
   \hline   
   \cite{chattopadhyay2015scalable} & Scalability & Both & RT, TR, NI, RE, AV  & Static & Heuristic & Deterministic & Local Search\\
   \hline    
   \cite{DBLPChattopadhyayB16} & Service Composition & Multiple & RT, TR, NI, RE, AV  & Static & Optimal + QoS constrained & Stochastic & ILP, A*, Local Search\\
                               &                     &          &                     &        & Heuristic                 &            & \\  
   \hline
   \cite{xia2013web,chen2012redundant} & Redundancy control & Single & RT, TR & Static & Optimal & Deterministic & Graph based search\\
   \hline  
   \cite{Zeng:2003:QDW:775152.775211} & Service Composition & Multiple & Price, Duration, RE, RP, AV & Static & Optimal & Deterministic & ILP\\
   \hline
   \cite{el2010tqos} & Transactional Driven & Multiple & Price, Duration & Static & Heuristic +  & Deterministic & Local optimization\\
                     & Service Composition  &          & SR, RP, AV      &        & Transactional Constraint &   & algorithm\\
   \hline
   \cite{deng2014top} & Top-k Service & Single & RT, TR & Static & Optimal  & Deterministic & Backtrack search, DFS\\
                      &  Composition  &        &         &        &          &               & parallel algorithm\\
   \hline
   \cite{oh2008effective} & Service Composition & Single & NI & Static & Optimal  & Deterministic & AI planning and \\
                      &    &      &        &             &          &                  & network optimization\\
   \hline
   \cite{yan2009qos} & Service Composition & Single & RT, TR & Static & Optimal & Deterministic & Graph based search\\
   \hline
   \cite{lecue2009optimizing} & Service Composition  & Single & RT, EP & Static & Optimal & Deterministic & Genetic Algorithms\\
   \hline
   \cite{paganelli2012qos} & Service Composition & Multiple & Cost, Duration, RP, RE, AV & Dynamic & Optimal & Deterministic & ILP\\
   \hline
   \cite{wagner2011qos} & Service Composition & Single & RT, TR & Static & Optimal & Deterministic & Graph search, planning algorithm\\
   \hline
   \cite{song2011workflow,pistore2005automated} & Service Composition & Multiple & Price, Duration, RP, ST, AV & Static & QoS constrained & Deterministic & Constraint Satisfaction Problem\\
   \hline
   \cite{yan2012anytime,yan2015anytime} & Service Composition & Multiple & RT, TR & Static & Heuristic & Deterministic & Anytime algorithm, Graphplan\\
   \hline
   \cite{wuqos} & Multi granularity & Multiple & Time, Price, RE & Static & Heuristic & Deterministic & Backtracking-based algorithm, \\
		& Service Composition     &        &         &        &          &                  & Genetic Algorithms\\
   \hline
   \cite{mostafa2015multi} & Service Composition in & Multiple & AV, RT, Cost & Dynamic & Optimal & Deterministic & Reinforcement learning, \\
			   & Uncertain Environments &          &              &         &         &               & Markov Decision Proces\\
   \hline
   \multicolumn{8}{l}{~}\\
   \multicolumn{8}{l}{Note: RT: ResponseTime, TR: Throughput, RE: Reliability, AV: Availability, ST: Success Rate, RP: Reputation, NI: Number of invocations
   EP: Execution Price}
  \end{tabular}\label{tab:relatedWork}
 \end{center}
\end{table*}

In contrast to existing literature, we propose an abstraction refinement based approach that aims to expedite the solution construction time by working on a reduced search space. Our approach provides a scalable way of pruning the dependency graphs that are considered by any composition solution. 
Our method has the ability to work on top of any service composition method, and improve its performance. Thus, we do not propose a new service composition solution, rather a framework on top of existing ones. This distinguishes our approach from the existing ones.



\section{Conclusion and future directions}
\label{sec:conclusion}
\noindent
This paper presents an abstraction-refinement based approach to expedite a service composition algorithm. 
For a large dataset, the abstraction can be very effective. It reduces the memory requirement and improves performance. As evident from the experimental
results, this mechanism is indeed more efficient on average, while having the same worst case performance when implemented on any method. Our method is generic enough to be 
applied to any QoS parameter in service composition. 
As future work, we are currently working on extending our proposal to develop more sophisticated refinement techniques and incorporating semantics based abstraction. We believe that our work will open up a lot of new research directions in the general paradigm of abstraction refinement based composition.

\scriptsize

\vspace{-0.45in}
\scriptsize
\begin{IEEEbiography}
[{\includegraphics[width=0.8in,height=0.8in,clip]{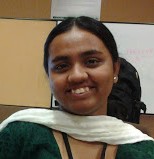}}]
{Soumi Chattopadhyay} is a Ph.D student at the Advanced Computing and Microelectronics Unit, Indian Statistical Institute Kolkata. 
She completed her under-graduate studies from West Bengal University of Technology, and Master's from the 
Indian Statistical Institute Kolkata -- all in Computer Science. 
Soumi's research interests are in distributed and services computing.
\end{IEEEbiography}
\vspace{-0.9in}

\begin{IEEEbiography}
[{\includegraphics[width=0.8in,height=0.8in,clip]{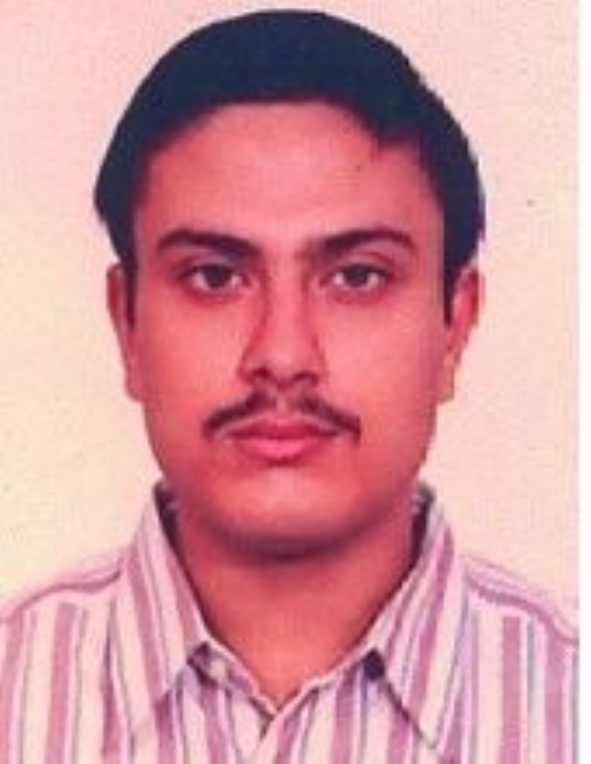}}]
{Ansuman Banerjee} is an Associate Professor at the Advanced Computing and
Microelectronics Unit, Indian Statistical Institute Kolkata. He received his
B.E. from Jadavpur University, and M.S. and Ph.D. degrees from the Indian
Institute of Technology Kharagpur -- all in Computer Science. His 
research interests include formal methods for services computing.
\end{IEEEbiography}

\end{document}